\documentclass[twoside,12pt]{article}
\usepackage{CJK}
\usepackage{indentfirst}
\usepackage{bm}
\usepackage{epsfig}
\usepackage{graphicx}
\usepackage{caption}
\usepackage{float}
\usepackage{subfig}
\usepackage{braket}
\usepackage{cite}
\usepackage{amsmath} 
\usepackage{amssymb} 
\usepackage{mdsymbol}
\usepackage{booktabs}
\usepackage[numbers,sort&compress]{natbib}

\usepackage{hyperref}

\begin{document}



\begin{center}
\LARGE\bf The Boundary Effect of QGP Droplet and Self-similarity Effect of Hadrons on QGP-hadron Phase Transition$^{*}$
\end{center}

\footnotetext{\hspace*{-.45cm}\footnotesize $^*$This work was supported by the National Natural Science Foundation of China under Grant No. 12175031, Guangdong Provincial Key Laboratory of Nuclear Science with No. 2019B121203010.}
\footnotetext{\hspace*{-.45cm}\footnotesize $^\dag$Corresponding author, E-mail: luancheng@dlut.edu.cn}

\begin{center}
\rm Tingting Dai$^{\rm a)}$, \ \ Huiqiang Ding$^{\rm a)}$,\ \ Luan Cheng$^{\rm a),b)\dagger}$, \\ Weining Zhang$^{\rm a)}$,\ \ and Enke Wang$^{\rm b)}$
\end{center}

\begin{center}
\begin{footnotesize} \sl
${}^{\rm a)}$ School of Physics, Dalian University of Technology, Dalian 116024, China\\
${}^{\rm b)}$ Institute of Quantum Matter, South China Normal University, Guangzhou 510631, China\\
\end{footnotesize}
\end{center}

\begin{center}
\footnotesize (Received XXXX; revised manuscript received XXXX)

\end{center} 

\vspace*{2mm}

\begin{center}
\begin{minipage}{15.5cm}
\parindent 20pt\footnotesize

We investigate the boundary effect of QGP droplet and self-similarity effect of hadrons on QGP-hadron phase transition. In intermediate or low energy collisions, when the transverse momentum is below QCD scale, QGP cannot be produced. However, if the transverse momentum fluctuates to a relatively large value, small scale QGP droplet is produced. The modified MIT bag model with multiple reflection expansion method is employed to study the QGP droplet with the curved boundary effect. It is found that the energy density, entropy density and pressure of QGP with the influence are smaller than those without the influence. In hadron phase, we propose Two-Body Fractal Model (TBFM) to study the self-similarity structure, arising from the resonance, quantum correlation and interaction effects. It is observed that energy density, entropy density and pressure increase due to the self-similarity structure. We calculate the transverse momentum spectra of pions with the self-similarity structure influence, showing a good agreement with the experimental data. Considering both the boundary effect and self-similarity structure influence, our model predicts an increase in the transition temperature compared to scenarios without these two effects in HIAF energy region $2.2\sim 4.5 \,\text{GeV}$.
\end{minipage}
\end{center}
\vspace*{2mm}
\begin{center}
\begin{minipage}{15.5cm}
\begin{minipage}[t]{2.3cm}{\bf Keywords:}\end{minipage}
\begin{minipage}[t]{13.1cm}
QGP droplet; multiple reflection expansion method; self-similarity structure; phase transition.
\end{minipage}\par\vglue8pt

\end{minipage}
\end{center}

\section{Introduction}
Quantum Chromodynamics (QCD) theory predicts that Quark-Gluon Plasma (QGP) can be formed after the de-confinement of quarks and gluons \cite{1, 2, 3}. In the past years, the production of QGP has been discovered in heavy-ion collisions\cite{4, 5,6}. To gain a clearer understanding of the properties of QGP, besides studying its properties, it is also significant to study QCD phase structure and QCD critical point \cite{7,8,9}. QCD phase diagram is an important tool to explore QCD phase structure and the critical point. Previous studies focused on exploring the phase diagram in the low chemical potential region\cite{8,10}. However, recent results from BES-II experiments indicate that the critical point may not exist in the high-energy collisions with the chemical potential $\mu_{B} < 450 \text{\ MeV}$\cite{11}. Consequently, it is significant to investigate the QCD critical point and the phase transition in mid-energy collisions with high chemical potential. At present, mid-energy experimental programs BES-I, BES-II at RHIC have successfully run, and the low and mid-energy facilities like HIAF, FAIR, and NICA have been proposed and are planned to run in the near future\cite{12,13}.

Previous studies concentrate on the phase structure in the thermodynamic limit\cite{14,15,16}. 
In the condition of high temperature or large baryon density, the production of large-scale QGP\cite{6, 17,18,19,20} can be approximately described in the thermodynamic limit\cite{21}. However, in the mid-energy collisions, where the baryon density is not large enough and the transverse momentum transfer fluctuates to the momentum less than QCD $\Lambda$ scale, QGP cannot be produced and only hadrons can be produced. While in the region where the transverse momentum transfer fluctuates to a higher momentum above QCD $\Lambda$ scale\cite{22}, small-scale QGP droplet can be produced\cite{23,24,25}. The curved boundary of these small-scale QGP droplets results in one side being QGP and the other side being hadron matter, inducing surface tension due to the differing densities\cite{26,27}. Hence, thermodynamic limit cannot be used in this  scenario, the effects of curved boundary and surface tension should be considered. In this paper, we will study the boundary effect on small-scale QGP and QGP-hadron transition.
After the phase transition, in the hadron phase, the collective flow as an important phenomenon observed in experiments\cite{28,29} results in the co-movement of the $\pi$ meson and its neighbouring $\pi$ meson. This will induce the resonance\cite{30, 31} in pion gas. The $\pi$-$\pi$ resonance effect for the pion gas will also be considered in our study.

In this paper, we will study the QGP-hadron phase transition with considering two influences: the surface tension and boundary effects for QGP droplets, and the $\pi$-$\pi$ resonance in pion gas. For the QGP droplet, the modified MIT bag model, extended by the multiple reflection expansion (MRE) method\cite{32}, is employed to describe the droplet with curvature and surface tension at the boundary. For pion gas, we propose the Two-Body Fractal Model (TBFM)\cite{33} to analyse the resonance influence on the pion gas near the critical temperature. Finally, we explore the QGP-hadron phase transition under the influences of the surface tension and resonance effects, and compare the result with that without these influences, demonstrating significant effects of the surface tension and resonance on the phase transition.

\section{Model and method}
\subsection{ The boundary effect of QGP droplet}

Under conditions of high temperature or large baryon density, a large amount of quarks and gluons are produced, leading to the formation of large-scale QGP\cite{34}. However in the mid energy collisions\cite{35,36,37}, where the temperature and density are not large enough, and the transverse momentum\cite{38,39} fluctuates to a small value less than QCD $\Lambda$ scale, free quarks and gluons cannot be produced, instead the produced are hadrons. Only in the region where the transverse momentum fluctuates to a large value larger than QCD $\Lambda$ scale, the small-scale QGP --- QGP droplet\cite{23,24}, can be produced. Consequently, for the QGP droplet, one side of the curved boundary is QGP, while the other side is hadrons. This will induce surface tension. The effects of the curved boundary and surface tension need to be considered.

MIT bag model is an effective model to describe the thermodynamics of quarks and gluons confined in a bag\cite{40}. The multiple reflection expansion method is an important approach to study the finite size effect of a system with the curved boundary in the powers of the system radius\cite{27,32,41}. It can be used to analyse the influence of curved boundary. In Madsen's work\cite{42,43}, MRE method used to be employed to investigate the effect of the curvature caused by the massive $s$ quark in the finite size MIT bag model. Here in our study, we apply MRE method to investigate the curvature of the boundary and the surface tension of the small-scale QGP droplet produced in the mid energy collisions. 

\begin{figure}[htbp]
\centering
\includegraphics[width=0.35\linewidth]{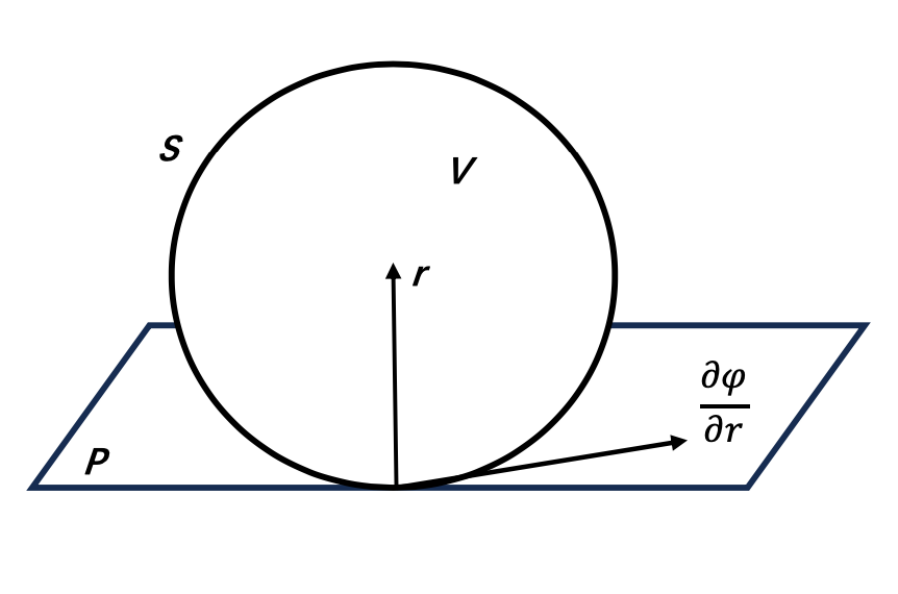}
\caption{The potential field of QGP droplet.}
\label{Fig01}
\end{figure}  

We treat the quarks and gluons inside the QGP droplet as a source field $\varphi$, the radius of the QGP droplet points to the droplet's center. 

The potential field distribution obeys the equation\cite{44}
\begin{equation}
\Delta \varphi+E \varphi=0.
\label{field}
\end{equation}
$\frac{\partial \varphi }{\partial r}$, recognized as the surface tension, is along the tangent plane $P$ of the point on the surface $S$ as shown in figure \ref{Fig01}.

In thermodynamic limit, the density of microstates with the momentum between $k$ and $k + dk$ is
\begin{equation}
    \rho(k) d k \sim \frac{V k^{2}}{2 \pi^{2}} d k.
    \label{rho0}
\end{equation}

However, the correction for the density of microstates in a small-scale QGP droplet due to the boundary effect needs to be considered. The correction can be calculated by using multiple reflection expansion for Green's function method.

The multiple reflection expansion for the time independent Green's function has the following form,
\begin{equation}
\begin{aligned}
S\left(\mathbf{r}, \mathbf{r}^{\prime}\right) & =S^{(0)}\left(\mathbf{r}, \mathbf{r}^{\prime}\right)+\oint_{\partial \Omega} d \sigma_{\alpha} S^{(0)}(\mathbf{r}, \boldsymbol{\alpha}) K(\boldsymbol{\alpha}) S^{(0)}\left(\boldsymbol{\alpha}, \mathbf{r}^{\prime}\right) \\
& +\oint_{\partial \Omega} d \sigma_{\alpha} d \sigma_{\beta} S^{(0)}(\mathbf{r}, \boldsymbol{\alpha}) K(\boldsymbol{\alpha}) S^{(0)}(\boldsymbol{\alpha}, \boldsymbol{\beta}) K(\boldsymbol{\beta}) S^{(0)}\left(\boldsymbol{\beta}, \mathbf{r}^{\prime}\right) \\
& +\cdots.
\label{mre}
\end{aligned}
\end{equation}
Here, $S^{(0)}$ corresponds to the free field satisfying $\left(i \gamma^{\mu} \partial_{\mu}-m\right) S^{(0)}\left(x-x^{\prime}\right)=\delta\left(x-x^{\prime}\right)$. $\Omega$ is the finite space of the QGP droplet. $K$ is the reflection kernel in the integral over the surface $\partial \Omega$, and describes the boundary effect using the corrections of increasing order in the principal radii of curvature as $(\frac{1}{R_1} + \frac{1}{R_2})^{n},\, n = 0, 1,\ldots$. Here, $R_1$ and $R_2$ are the principal radii of curvature, $R_{1} = R_{2} =R$. $\mathbf{r}$ and $\mathbf{r}^{\prime}$ represent the spatial vectors in $\Omega$. $\boldsymbol{\alpha}$ and $\boldsymbol{\beta}$ are the spatial vectors on $\partial \Omega$. 

Based on the relation\cite{32} of the density of microstates $\rho(\omega)$ and $S$ matrix,
\begin{equation}
    \rho(\omega)  =\mp \frac{1}{\pi} \operatorname{Im} \int_{\Omega} d^{3}r \operatorname{tr} [S(\mathbf{r}, \mathbf{r}^{\prime},\omega \pm i \varepsilon) \gamma^{0}].
\label{rho1}
\end{equation}

We can obtain the multiple reflection expansion of density of states in powers of $\frac{1}{kR}$, where $R$ is the radius of QGP droplet, 
\begin{equation}
    \rho(k)=\frac{V k^{2}}{2 \pi^{2}}+f_{S}\left(\frac{m}{k}\right) k A+f_{C}\left(\frac{m}{k}\right) C+\ldots.
\label{mrerho}
\end{equation}
The second term in equation (\ref{mrerho}) corresponds to the $n = 0$ order for boundary curvature correction. Integrating the kernel $K$ over the surface, $A$ can be derived as the surface area of QGP droplet, $A = 4\pi R^{2}$. The third term in equation (\ref{mrerho}) arises from the curvature of the boundary with the $n=1$ order, $C$ represents the extrinsic curvature given by the surface integral $C =\oint_{\partial \Omega} d^{2} \sigma (\frac{1}{R_1(\sigma)} + \frac{1}{R_2(\sigma)}) = 8\pi R$. Considering different corrections of orders $n =0,1$ in the principal radii of curvature within the second term of the Green function equation (\ref{mre}), $f_{S}$ and $f_{C}$ are derived as
\begin{equation}
    f_{S} = - \frac{1}{8 \pi}[1 - \frac{2}{\pi} \arctan(\frac{k}{m})],
\label{fs}
\end{equation}
\begin{equation}
     f_{C} =\frac{1}{12 \pi^2}\{1 - \frac{3 k }{2 m}[\frac{\pi}{2}- \arctan(\frac{k}{m})]\}.
\label{fc}
\end{equation}
Taking the surface tension and curvature of QGP droplets into account, the density of microstates for quarks and anti-quarks with momentum $p_{q}$ is obtained from equation (\ref{mrerho}) as
\begin{equation}
\rho_{qMRE} = \rho_{\bar{q}MRE} =g_{q} V p_{q}^{2}[\frac{1}{2\pi^{2}} + \frac{3}{p_{q}R} f_{S} + \frac{6p_{q}^{2}}{(p_{q}R)^{2}}f_{C}].
\label{qMRE}
\end{equation}
For gluons with the mass $m=0$, it follows that $f_{s} =0$, $f_{C} = -\frac{1}{6\pi^{2}}$. The density of microstates is
\begin{equation}
\rho_{gMRE} = g_{g}(\frac{V p_{g}^{2}}{2 \pi^{2}} - \frac{V}{\pi^{2} R^{2}}),
\label{gMRE}
\end{equation}
where $p_{g}$ is the momentum of gluons. $g_{q}$ and $g_{g}$ denote the degeneracy factor of the quarks (anti-quarks) and gluons. 

Applying the density of microstates under the influence of the surface tension and the curvature above, the pressure of QGP droplet is expressed as
\begin{equation}
P_{QGP} = - \int_{0}^{\infty} \rho_{qMRE} f_{q} \frac{\partial \varepsilon_{q}}{\partial V} \,dp_{q} - \int_{0}^{\infty} \rho_{\bar{q}MRE} f_{\bar{q}} \frac{\partial \varepsilon_{\bar{q}}}{\partial V}\,dp_{q} -\int _{0} ^{\infty} \rho_{gMRE} f_{g} \frac{\partial \varepsilon_{g}}{\partial V}\, dp_{g} - B(T, \mu),
\label{pinbag}
\end{equation}
where $\varepsilon_{q(\bar{q})}, \varepsilon_{g}$ are the energy of the quarks (anti-quarks) and gluons, $\varepsilon_{q} = \varepsilon_{\bar{q}} = \sqrt{m^{2} + p^{2}}$, $\varepsilon_{g} = p_{g}$.  $f_{q}, f_{\bar{q}}$ denote the Fermi distribution for quarks and anti-quarks  $f_{q} = f_{\bar{q}}$, $f_{g}$ is the Bose distribution of gluons. The three terms in equation (\ref{pinbag}) correspond to quark, anti-quark and gluon respectively. $B(T,\mu)$ is the bag pressure as a function of the temperature $T$ and chemical potential $\mu$ in MIT bag model\cite{45,46}. To investigate the influence of boundary effect, we adjust the parameters to align with the recent result of lattice QCD\cite{47,48}, obtaining the bag pressure $B(T,\mu)$ as
\begin{equation}
    B(T, \mu) = B_{0} - ( \frac{\mu^{4}}{85\pi^{2}} + \frac{1}{4}T^{2}\mu^{2} + \frac{1}{10} \pi^{2} T^{4}),
    \label{B}
\end{equation}
where $B_{0}^{1/4} = 233.5\, \text{MeV}$.

We also obtain the energy density and the entropy density of the QGP droplet under the influence of surface tension and curvature boundary as the following,
\begin{equation}
 u_{QGP} = \frac{1}{V}\int_{0}^{\infty}\rho_{qMRE} f_{q} \varepsilon_{q} \, dp_{q} + \frac{1}{V}\int_{0}^{\infty}\rho_{\bar{q}MRE} f_{\bar{q}} \varepsilon_{\bar{q}} \, dp_{q}  +\frac{1}{V}\int_{0}^{\infty}\rho_{gMRE} f_{g} \varepsilon_{g} \,dp_{q} ,
\end{equation}
 \begin{equation}
 \begin{aligned}
 s_{QGP}  &= -\frac{1}{V}\int_{0}^{\infty} \rho_{qMRE} [f_{q} ln f_{q} - (1+f_{q}) ln (1 + f_{q}) ] \,dp_{q}-\frac{1}{V} \int_{0}^{\infty} \rho_{\bar{q}MRE} [f_{\bar{q}} ln f_{\bar{q}}  \\
& - (1+f_{\bar{q}}) ln (1 + f_{\bar{q}}) ] \,dp_{q}-\frac{1} {V}\int _{0} ^{\infty} \rho_{gMRE} f_{g} [f_{g} ln f_{g} - (1+f_{g}) ln (1 + f_{g})] \, dp_{g}.
\end{aligned}
\end{equation}

\begin{figure}[ht]
\centering

\subfloat[\label{Fig02a} Energy density]{\includegraphics[width= 0.5\linewidth]{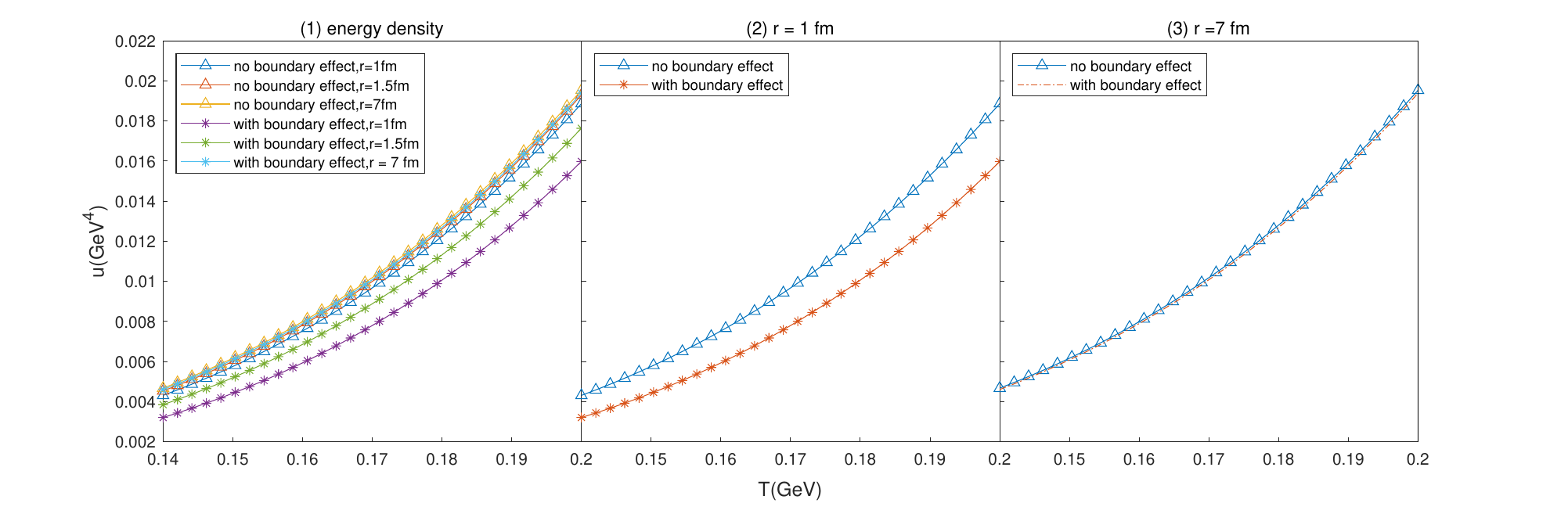}}

\subfloat[\label{Fig02b}Entropy density]{\includegraphics[width=0.5\linewidth]{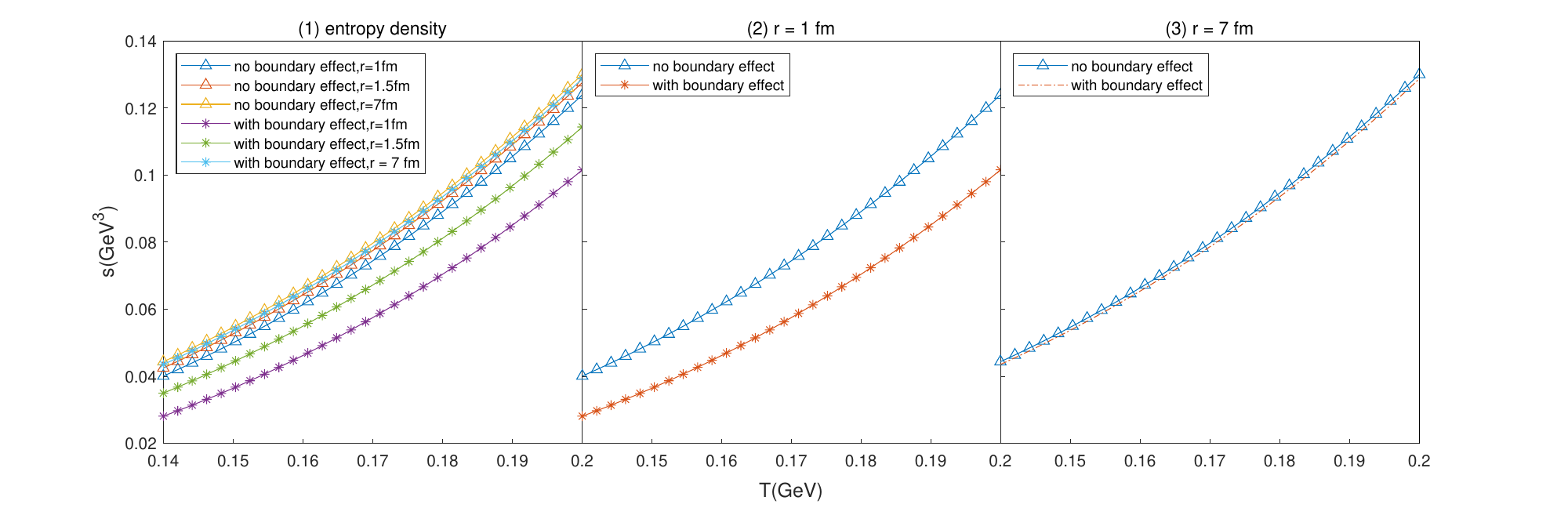}}

\subfloat[\label{Fig02c}Pressure]{\includegraphics[width=0.5\linewidth]{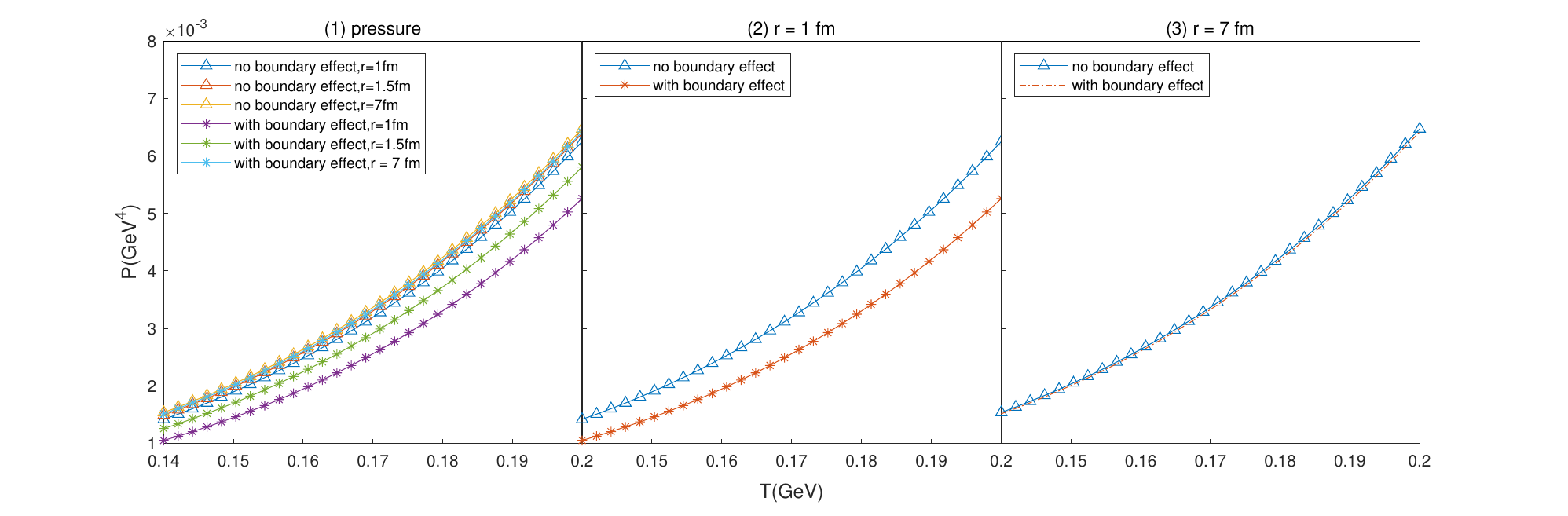}}
\caption{Thermodynamic quantities of QGP droplets with radius $r = 1, 1.5, 7\text{\ fm}$ respectively.}
\label{Fig02}
\end{figure}
The energy density, entropy density, and the pressure of QGP droplets with and without boundary effect as functions of temperature at fixed radius $r = 1 \text{\ fm},1.5 \text{\ fm},7\text{\ fm}$ are depicted in figure \ref{Fig02}. The energy density, entropy density, and the pressure increase with the temperature as shown in figure \ref{Fig02a}(1), figure \ref{Fig02b}(1), figure \ref{Fig02c}(1), respectively. For the sake of analysis, we compare the energy density, entropy density, and the pressure with (red trace) and without (blue trace) the boundary effect at fixed QGP radius $r = 1 \text{\ fm}$ in figure \ref{Fig02a}(2), figure \ref{Fig02b}(2), figure \ref{Fig02c}(2). It is found that at smaller radius $r = 1 \text{\ fm}$, three physical properties under the influence of boundary effect are less than those without the influence. This demonstrates that the boundary effect decreases the three physical properties at small QGP size $r = 1 \text{\ fm}$. We also compare the energy density, entropy density, and the pressure with (red trace) and without (blue trace) the boundary effect at fixed QGP radius $r = 7 \text{\ fm}$ as shown in figure \ref{Fig02a}(3), figure \ref{Fig02b}(3), figure \ref{Fig02c}(3), respectively. The three physical properties with and without the boundary effect are approximately the same, suggesting that the influence of boundary effect is significant at QGP size, when the size increases, the influence decreases.

In summary, the surface tension originating from finite size effect of the small-scale QGP droplets impacts on the thermodynamics of QGP droplets, especially in the mid and low collisions where the small QGP droplets can be produced.
 
\subsection{Two-body structure in pion gas}\label{sec2.2}
\subsubsection{Two-body Fractal Model}
\begin{figure}[htbp]
\centering
\subfloat[quark aspect]{\includegraphics[width = 0.25\linewidth]{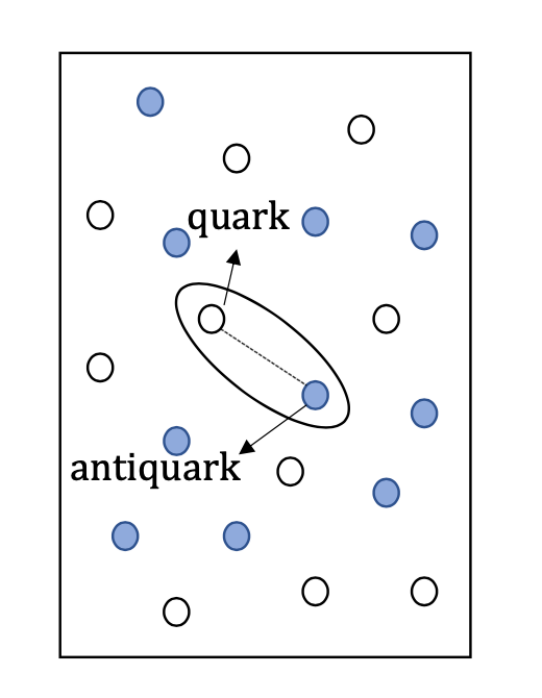}\label{Fig03a} }
\subfloat[meson aspect]{\includegraphics[width = 0.25\linewidth]{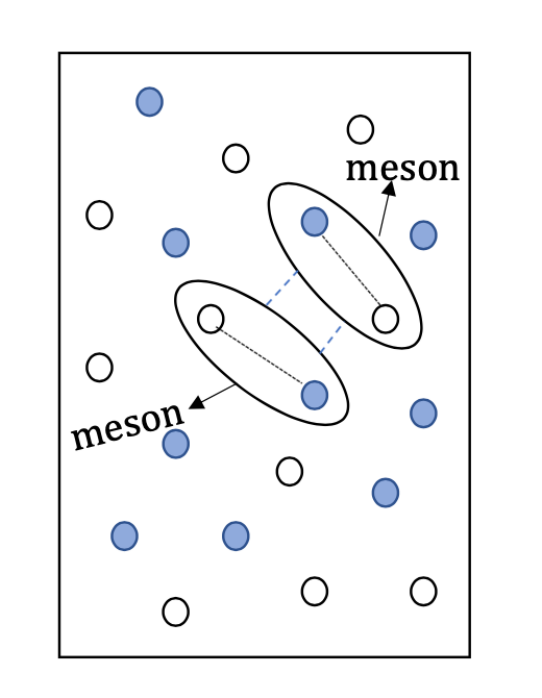}\label{Fig03b}}
\caption{Self-similarity structure of a meson and a resonant state.}
\label{Fig03}
\end{figure}

The collective flow is a significant phenomenon in collision experiments. Due to the collective flow, $\pi$ mesons have similar momentum, frequency with the neighbours, resulting in co-movement of the neighbouring $\pi$ mesons. Furthermore, the same frequency of the neighboring $\pi$ mesons will surely induce resonance\cite{49}. Secondly, in high energy collisions, the pion density at the phase transition temperature is approximately $ \frac{N}{V}=0.5 \text{\ fm}^{-3}$, the distance between the neighbouring pions is approximately $(\frac{V}{N})^{1/3} \approx 1.3\text{\ fm}$, which is less than the thermal length of a $\pi$ meson $\lambda_{T} = \frac{h}{\sqrt{2 \pi m k T}}  = 3.008 \text{\ fm}$. This implies the quantum correlation between two neighbouring $\pi$ mesons. In mid or low energy collisions with larger chemical potential and larger baryon density, the distance between the neighboring mesons is less than that in the high energy collisions. Thus, the quantum correlation is larger than that case. Thirdly, the solution to the Two-Body Dirac Equation for $u\bar{d}$ quarks\cite{50} suggests that within a distance less than $l=1.2 \text{\ fm}$, a $u$ quark has a strong interaction with $\bar{d}$ quark. Meanwhile, the interaction distance and the pion average distance satisfy $(\frac{V}{N})^{1/3}<2l$, illustrating the strong interaction between a $\pi$ meson and its nearest neighbour. Overall, under the influence of resonance, quantum correlation and strong interaction in high baryon density region, the $\pi$ meson and the nearest neighbour can form the two-body $\pi$-$\pi$ system as depicted in figure \ref{Fig03b}.

Analogy to $\pi$-$\pi$ resonate state as a two-body structure, a $\pi$ meson is also recognized as a two-quark bound state as shown in figure \ref{Fig03a}. Owing to the two-body structure, we postulate that the two-quark system and two-meson system exhibit the self-similarity and apply the fractal theory to describe the characteristic. The concept of fractal was originally proposed by Mandelbort\cite{51,52} to describe the similarity between the components of a structure and the structure as a whole. Tsallis statistics, inspired by the fractal theory, has been widely applied to study systems characterized by self-similarity fractal structures\cite{53,54,55} in different scales. Hence, we will propose the Two-Body Fractal Model (TBFM)\cite{33} which is based on the fractal inspired Tsallis statistics to analyse a $\pi$ meson and $\pi$-$\pi$ bound state as the self-similarity structure in pion gas.

\subsubsection{Pion gas influenced by self-similarity structures}
Firstly, from meson aspect, we study the $\pi$ meson as a two-quark bound state. The $\pi$ meson is influenced by the self-similarity structure from the resonance, quantum correlation, and strong interaction. We introduce the modification factor $q_{1}$ to denote the self-similarity structure influence on the $\pi$ meson. In the rest frame, the probability of the $\pi$ meson is
\begin{equation}
P _{\pi 1} = \frac{\bra{\psi_{0}}[1 + (q_{1} -1)\beta \hat{H}]^{\frac{q_{1}}{1  - q_{1}}}\ket{\psi_{0}}} {\sum_{i} \bra{\psi_{i}}[1 + (q_{1} -1)\beta \hat{H}]^{\frac{q_{1}}{1  - q_{1}}}\ket{\psi_{i}}},
\label{p1}
\end{equation}
in which $\beta$ is the inverse of temperature, $\beta = 1/T$. $\psi_{0}$ is the wavefunction of a $\pi$ meson, which is the ground state of $u\bar{d}$ bound state. $\psi_{i}$ is that of the $u \bar{d}$ bound states in different eigenstates. The Hamiltonian of $u\bar{d}$ bound states is
\begin{equation}
   \hat{H} = \sqrt{m_{q}^2 + \hat{p_1}^2} +\sqrt{m_{q}^2 + \hat{p_2}^2} + \hat{V}(r), 
\label{H}
\end{equation}
where $\sqrt{m_{q}^2 + \hat{p_{1(2)}}^{2}}$ is the kinetic energy of $u,\bar{d}$ quarks, $m_{q} = 55 MeV$ is the corresponding mass. $\hat{V}(r)$ is the quark potential. Here, we consider the quark potential $V(r)$ from the work of calculating Two Body Dirac Equation (TBDE) done by Crater group\cite{50} as 
\begin{equation}
V(r)=\frac{8\pi \Lambda^{2} r}{27} - \frac{16 \pi}{27 r \ln [(K e^{2} + B_{1}/(\Lambda r)^{2}]},
\label{interaction}
\end{equation}
where $\Lambda = 0.4218 \text{\  GeV}$ is the scale size,  $K = 4.198$ and $B_{1}=0.05081$ are the variable parameters, $e$ is Euler's constant. 

The denominator in equation (\ref{p1}) represents the partition function, which is the sum of the probability in all microstates, 
\begin{equation}
\begin{aligned}
&\sum_{i} \bra{\psi_{i}}[1 + (q_{1} -1)\beta \hat{H}]^{\frac{q_{1}}{1  - q_{1}}}\ket{\psi_{i}}= [1 + (q_{1} -1)\beta E_{\pi}]^{\frac{q_{1}}{1  - q_{1}}} + [1 + (q_{1} -1)\beta E_{\rho}]^{\frac{q_{1}}{1  - q_{1}}} \\
&+ \frac{V}{(2 \pi)^{6}}\int_{p_{\text{min}}} ^{\infty } \int_{r_{\text{min}}} ^{r_{\text{max}}} [1 + (q_{1} -1)\beta (\sqrt{m_{q}^2 + \hat{p_1}^2} +\sqrt{m_{q}^2 + \hat{p_2}^2} + \hat{V}(r))]^{\frac{q_{1}}{1  - q_{1}}} 4\pi r^{2} \, dr\, d^{3}\vec{p_1}\, d^{3}\vec{p_2}.
\label{partition}
\end{aligned}
\end{equation}
For the calculation convenience, we sum up the lower discrete energy levels of $\pi$ and $\rho$ which are measured in the experiments in the first and second terms in equation (\ref{partition}), and integrate nearly continuous higher energy levels in the third term. In the first and second terms, for the energy of $\pi$ and $\rho$,  $E_{\pi}= 0.159 \text{\  GeV}$, $E_{\rho} = 0.792 \text{\ GeV}$. In the third term, $V$ is the motion volume of $\pi$ meson, we take $V$ as the volume of hadron gas, $V = \frac{4}{3}\pi r_{0}^{3}$, $r_{0}$ is the radius of the hadron gas. $r_{\text{min}}$ is the lower limit of the quark distance, with its value set as the lattice spacing for light quarks\cite{56,57} shown in table \ref{TableI}. $r_{\text{max}}$ is the upper limit of quark distance, here we set $r_{\text{max}} = r_{0}$. $p_{\text{min}} = 0.29 \text{\ GeV}$ is the lower limit of quark momentum as the quark momentum at $\rho$ state.

In the above, we have discussed the escort probability of $\pi$ meson while considering self-similarity structure based on the fractal theory. Tsallis entropy is associated  with the escort probability in multifractal and follows the maximum entropy principle\cite{55, 58,59}. Meanwhile, entropy plays a crucial role in studying physical properties. In our model, we derive the Tsallis entropy of $\pi$ meson as 
\begin{equation}
S_{\pi 1}=\frac{1-\sum_{i=1}^{W} P_{1 i}^{q_{1}}}{q_{1}-1} = \frac{1}{q_{1} -1} \{1 - \frac{\sum_{i} \bra{\psi_{i}}[1 + (q_{1} -1)\beta \hat{H}]^{\frac{q_{1}}{1  - q_{1}}}\ket{\psi_{i}} }{\{\sum_{i} \bra{\psi_{i}}[1 + (q_{1} -1)\beta \hat{H}]^{\frac{1}{1  - q_{1}}}\ket{\psi_{i}} \}^{q_{1}}}\}.
\label{s1}
\end{equation}

Secondly, we analyse a $\pi$ meson influenced by self-similarity structure in quark aspect. We introduce the modification factor $q_{0}$ of the self-similarity structure influence on $u$ or $\bar{d}$ quarks, which comes from the influence of the strong interaction between $u$ and $\bar{d}$ inside $\pi$ meson, and the influence of outside hadrons on pions. The probability of the $u$,$\bar{d}$ quarks inside $\pi$ meson also obeys the power-law form as
\begin{equation}
P_{u} = P _{\bar{d}} = \frac{\bra{\phi_{q0}}[1 + (q_{0} -1)\beta \hat{H_q}]^{\frac{q_{0}}{1  - q_{0}}}\ket{\phi_{q0}}}{\sum_{j} \bra{\phi_{qj}}[ 1 + (q_{0} -1)\beta \hat{H_{q}}]^{\frac{q_{0}}{1  - q_{0}}}\ket{\phi_{qj}}},
\label{pquark}
\end{equation}
where $\hat{H_q}$ is the Hamiltonian of quark and anti-quark with $\hat{H_q} =\hat{H}_{\bar{q}} = \sqrt{m_{q}^{2} + \hat{p}^{2}}$. $\phi_{q0}$ is the wavefunction of the $u, \bar{d}$ quarks at the $\pi$ state, $\phi_{qj}$ is the wavefunction of the $u, \bar{d}$ quark at different bound states. 

The escort probability of $\pi$ meson can be expressed as the product of the escort probability of $u$ and $\bar{d}$ quarks, $P_{\pi 2} = P_{u} \cdot P_{\bar{d}}$. Because of the pseudoadditivity\cite{53} in non-extensive statistics, we define the escort parameter $q_{2}$ and it satisfies
\begin{equation}
 \bra{\phi_{j}}[1 + (q_{2} - 1) \beta \hat{H_{t}} ]^{\frac{q_{2}}{1 - q_{2}}}\ket{\phi_{j}} = \bra{\phi_{qj}} [ 1 + (q_{0} -1)\beta \hat{H_{q}} ]^{\frac{ q_{0}}{1- q_{0}}}\ket{\phi_{qj}} \cdot \bra{\phi_{qj}} [ 1 + (q_{0} -1)\beta \hat{H_{q}} ]^{\frac{ q_{0}}{1- q_{0}}}\ket{\phi_{qj}},
\end{equation}
where $\hat{H_{t}} = \sqrt{m_{q}^{2} + \hat{p_1}^{2}} + \sqrt{m_{q}^{2} + \hat{p_2}^{2}}$, $\phi_{j}$ is the wavefunction of the $u\bar{d}$ bound system.

So the escort probability of $\pi$ meson is given by
\begin{equation}
P_{\pi 2} = P_{u} \cdot P_{\bar{d}}=\frac{\bra{\phi_{0}}[1 + (q_{2} -1)\beta \hat{H_t}]^{\frac{q_{2}}{1  - q_{2}}}\ket{\phi_{0}}}{\sum_{j} \bra{\phi_{j}}[ 1 + (q_{2} -1)\beta \hat{H_{t}}]^{\frac{q_{2}}{1  - q_{2}}}\ket{\phi_{j}}},
\label{p2}
\end{equation}
where $\phi_{0}$ is the wavefunction of $u\bar{d}$ bound system at $\pi$ bound state. The denominator in equation (\ref{p2}) is the partition function for the quark and it sums up the probability of  $u, \bar{d}$ quarks in all microstates as
\begin{equation}
\begin{aligned}
&\sum_{j} \bra{\phi_{j}}[ 1 + (q_{2} -1)\beta \hat{H_{t}}]^{\frac{q_{2}}{1  - q_{2}}}\ket{\phi_{j}} = [1 + (q_{2} -1)\beta E_{u\bar{d}\pi}]^{\frac{q_{2}}{1  - q_{2}}} + [1 + (q_{2} -1)\beta E_{u\bar{d}\rho}]^{\frac{q_{2}}{1  - q_{2}}} \\
& + \frac{V_{q}^{2}}{(2 \pi)^{6}}\int_{p_{\text{min}}} ^{\infty } [1 + (q_{2} -1)\beta (\sqrt{m_{q}^{2} + \hat{p_1}^{2}} + \sqrt{m_{q}^{2} + \hat{p_2}^{2}})]^{\frac{q_{2}}{1  - q_{2}}} d^{3}\vec{p_1}d^{3}\vec{p_2},
\end{aligned}
\end{equation}
where $E_{u\bar{d}\pi},E_{u\bar{d}\rho}$ represent the kinetic energy of $u\bar{d}$ quarks at $\pi$ and $\rho$ state respectively, $E_{u\bar{d}\pi} = 0.8508 \text{\ GeV}, E_{u\bar{d}\rho} = 0.3085 \text{\ GeV}$. $V_{q}$ denotes the motion volume of $u, \bar{d}$ quarks, which approximately equals the motion volume of $\pi$ meson $V$.

Meanwhile, the corresponding Tsallis entropy of $\pi$ meson is given by
\begin{equation}
S_{\pi 2}=\frac{1-\sum_{j=1}^{W} P_{2 j}^{q_{2}}}{q_{2}-1} = \frac{1}{q_{2}-1} \{ 1 - \frac{\sum_{j} \bra{\phi_{j}}[ 1 + (q_{2} -1)\beta \hat{H_{t}}]^{\frac{q_{2}}{1  - q_{2}}}\ket{\phi_{j}}}{\{\sum_{j} \bra{\phi_{j}}[ 1 + (q_{2} -1)\beta \hat{H_{t}}]^{\frac{1}{1  - q_{2}}}\ket{\phi_{j}}\}^{q_{2}}}\}.
\label{s2}
\end{equation}

Overall, we have analyzed the $\pi$ meson from meson and quark aspects respectively. In the meson aspect, the $\pi$ meson satisfies self-similarity. By introducing the modification factor $q_1$, we obtain the probability of $\pi$ meson as in equation (\ref{p1}) and its entropy as in equation (\ref{s1}). In the quark aspect, the $u, \bar{d}$ quarks also satisfy the self-similarity. Their probabilities follow the power-law form, leading to the probability of $\pi$ meson being the product of the probabilities of $u, \bar{d}$ quarks. The escort parameter $q_{2}$ is applied to derive the probability of  $\pi$ meson as in equation (\ref{p2}) and entropy as in equation (\ref{s2}). In fractal theory, the probabilities and Tsallis entropies for $\pi$ meson in two aspects should equal each other, thus
\begin{equation}
P_{\pi 1} = P_{\pi 2}, \qquad  S_{\pi 1} = S_{\pi 2}.
\label{sets}
\end{equation}

Considering different situations at different collision energies, the conservation equation sets (\ref{sets}) can be applied to evaluate the two-body self-similarity structure influence on the pion gas at different collision energies. The parameter $q_{1}$ describes the self-similarity structure influence on the $\pi$ meson in meson aspect, and the escort parameter $q_{2}$ denotes the quark interaction in the $\pi$ meson and the self-similarity structure influence on the $\pi$ meson in the quark aspect. 
\begin{figure}[ht]
\centering
\includegraphics[scale=0.4]{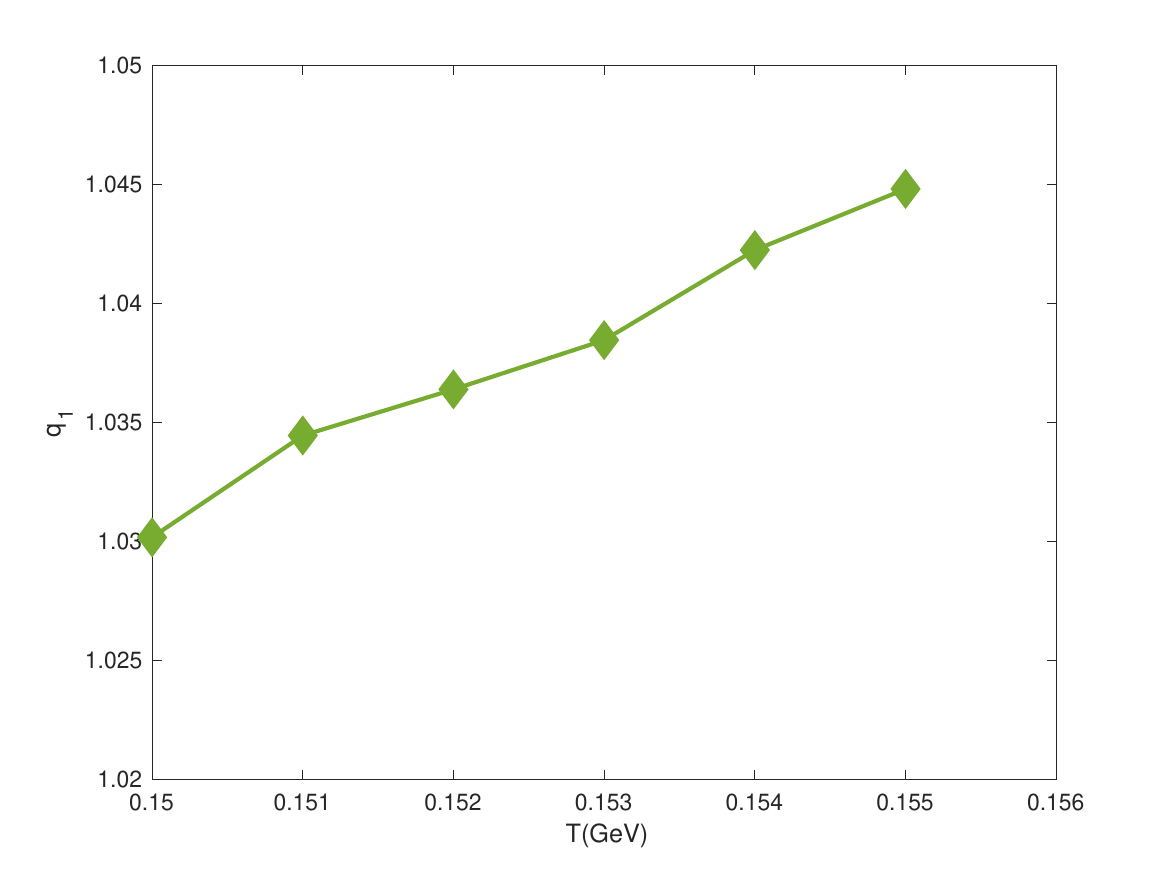}
\caption{The relationship between the influence factor $q_{1}$ and the temperature $T$ near to the critical temperature in Au+Au collision at $\sqrt{s_{\text{NN}}} =  39\,\text{GeV}$ .}
\label{Fig04}
\end{figure}

To study the temperature evolution of the modification factor $q_{1}$, we solve the equation sets (\ref{sets}) to obtain $q_{1}$ in Au+Au collision at $\sqrt{s_{\text{NN}}} =  39 \,\text{GeV}$. Shown in figure \ref{Fig04} is the modification factor $q_{1}$ at $\sqrt{s_{\text{NN}}} =  39 \,\text{GeV}$. It is found that $q_{1} >1$, demonstrating that the similarity structure decreases the number of microstates. In the context of non-extensive statistics, if $q_{1}>1$, $S_{q} < S_{B-G}$. Moreover, it is found that $q_{1}$ decreases with decreasing temperature. So that the pion gas is typically influenced by the self-similarity structure near to the phase transition temperature.

From the lattice QCD phase diagram\cite{8}, we extract the temperature and chemical potential of the phase transition within the range of error for different collision energies, which is listed in the second column of table \ref{TableI}. Substituting the temperature and the radius of motion volumes, $q_1$, $q_2$ can be solved in equation sets (\ref{sets}). Table \ref{TableI} presents the parameters $q_1, q_2$ solved for Au+Au collisions at $\sqrt{s_{\text{NN}}} = 7.7, 11.5, 19.6, 27, 39 \,\text{GeV}$ for $0-5\%$ centrality. $q_1$ is larger than 1 in all collision energies, suggesting the existence of the self-similarity structure. As the colliding energy increases, the temperature of phase transition and the motion volume of $\pi$ mesons increases, $q_1$ also becomes larger. The tendency suggests that the influence of $\pi-\pi$ self-similarity structure becomes stronger with the increase of the colliding energy. This is consistent with that at larger collision energies, the collective flow effect is stronger, leading to stronger resonance, so the self-similarity structure influence becomes stronger.

\begin{table}[htbp]
\caption{ The factors $q_{1}$ and $q_{2}$ in Au+Au collisions at $\sqrt{s_{\text{NN}}} = 7.7, 11.5, 19.6, 27, 39 \text{\ GeV} $ for  $ 0-5\% $ centrality solved by TBFM.}
\label{TableI}
\resizebox{\textwidth}{!}{
\begin{tabular}{ccccccc}
\hline
{$\sqrt{s_{\text{NN}}}/\text{\ GeV}$}&{ $T / \text{\,GeV}$ }&{$\mu_{B} / \text{\ GeV}$}&{$r_{\text{min}}/\text{\ fm}$}&{ $r_{0}/\text{\ fm}$}&{ $q_{1}$ }&{$q_{2}$} \\
 \hline
 {7.7}&{0.1424 $\pm$ 0.00137} & {0.42} &{0.14} & {6.7} & {1.03062 $\pm$ 0.001100}& {1.13426 $\pm$ 0.004055} \\
 \hline
 {11.5} & {0.1483 $\pm$ 0.00142} & {0.316}&{0.12} & {6.7}& {1.03451 $\pm$ 0.002688} & {1.12593 $\pm$ 0.006648} \\
\hline
{19.6} & {0.1527 $\pm$ 0.00147} & {0.206} &{0.11}& {6.95} & {1.03595 $\pm$ 0.002156}& {1.11965 $\pm$ 0.006417}\\
\hline
{27} & {0.1541 $\pm$ 0.00148} & {0.156} &{0.11}& {6.95} & {1.04224 $\pm$ 0.002919}& {1.10896 $\pm$ 0.005396 }  \\
\hline
{39} & {0.155 $\pm$ 0.00149} & {0.112}&{0.11} & {7} & {1.04482 $\pm$ 0.001929}& {1.10500 $\pm$ 0.004225}  \\
\hline
\end{tabular}}
\end{table}

\begin{figure}[htbp]
\centering
\includegraphics[scale=0.4]{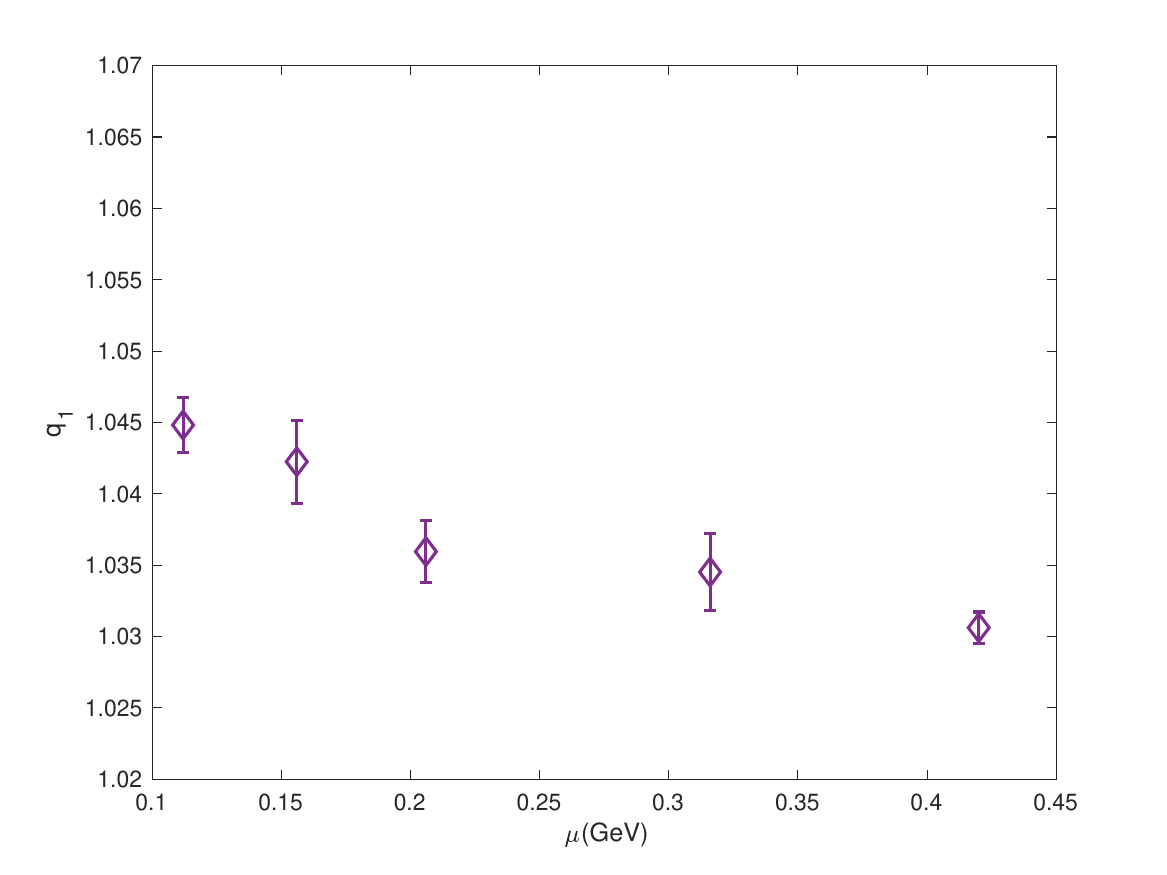}
\caption{The relationship between the modification factor $q_{1}$ and baryon chemical potential $\mu$.}
\label{Fig05}
\end{figure}

We analyse the values of chemical potential and self-similarity modification factor $q_{1}$ as shown in figure \ref{Fig05}, and find that with the decrease of the collision energy, the chemical potential $\mu$ increases, the modification factor $q_{1}$ decreases. The variation of $q_{1}$ with $\mu$ follows a specific rule, as summarized in the parameter equation
\begin{equation}
    q_{1} = 0.0393 e ^{- 1.9342\mu} + 1.0132.
    \label{fit}
\end{equation} 
Based on this rule, in the planned collision energy region $\sqrt{s_{\text{NN}}} = 2.2 \sim 4.5 \text{\ GeV}$ in HIAF, we obtain $q_{1} = 1.0208 \sim 1.0224 $.

Regarding the $u\bar{d}$ bound system as a grand canonical ensemble and based on the escort probability of pion in equation (\ref{p1}), the normalized density operator $\hat{\rho}$ is expressed as\cite{60} 
\begin{equation}
    \hat{\rho}=\frac{\left[1+\left(q_{1}-1\right) \beta(\hat{H}-\mu \hat{N})\right]^{\frac{q_{1}} {\left(1-q_{1}\right)}}}{\left.\operatorname{Tr}\left[1+\left(q_{1}-1\right) \beta(\hat{H}-\mu \hat{N})\right]\right]^{\frac{q_{1}} {\left(1-q_{1}\right)}}},
\end{equation}
where $\hat{N}$ denotes the particle number operator of the grand canonical ensemble. 

Based on the density operator $\hat{\rho}$ and the pseudoadditivity law\cite{61,62,63}, the particle number distribution for $\pi$ meson can be obtained as\cite{33}
\begin{equation}
    \bar{n}_{\pi} =\frac{1}{\left[1+\left(q_{1}-1\right) \beta\left(\varepsilon_{\pi}-\mu\right)\right]^{\frac{q_{1}} {\left(q_{1}-1\right)}}-1},
\end{equation}
where $\varepsilon_{\pi}$ is the energy of $\pi$ meson. In the limit $q_{1} \rightarrow 1$, the distribution converges to become Bose-Einstein distribution.

With the $\pi$ meson distribution, the energy density $u_{\pi}$, pressure $P_{\pi}$ and entropy density $s_{\pi}$ can be calculated through the following formula
\begin{equation}
u_{\pi}=\frac{g_{\pi}}{2\pi^{2}} \int_{0}^{\infty} \varepsilon_{\pi}  \bar{n}_{\pi} p_{\pi}^{2} d p_{\pi},
\end{equation}
\begin{equation}
P_{\pi}=- \frac{g_{\pi}}{2\pi^{2}}\int_{0}^{\infty}  \bar{n}_{\pi} \frac{\partial \varepsilon_{\pi}}{\partial V} p_{\pi}^{2} d p_{\pi},
\label{pionp}
\end{equation}
\begin{equation}
s_{\pi}=-\frac{g_{\pi}}{2\pi^{2}} \int_{0}^{\infty} [\bar{n}_{\pi} \ln  \bar{n}_{\pi}-\left(1+\bar{n}_{\pi}\right) \ln \left(1+\bar{n}_{\pi}\right)] p_{\pi}^{2}d p_{\pi},
\end{equation}
where $p_{\pi}$ is the momentum of $\pi$ meson. $g_{\pi} = 3$ is the degeneracy of $\pi$ meson and $V$ is the volume of pion gas.

\begin{figure}[htbp] 
\subfloat [Energy density]{\includegraphics[width= 0.3\linewidth]{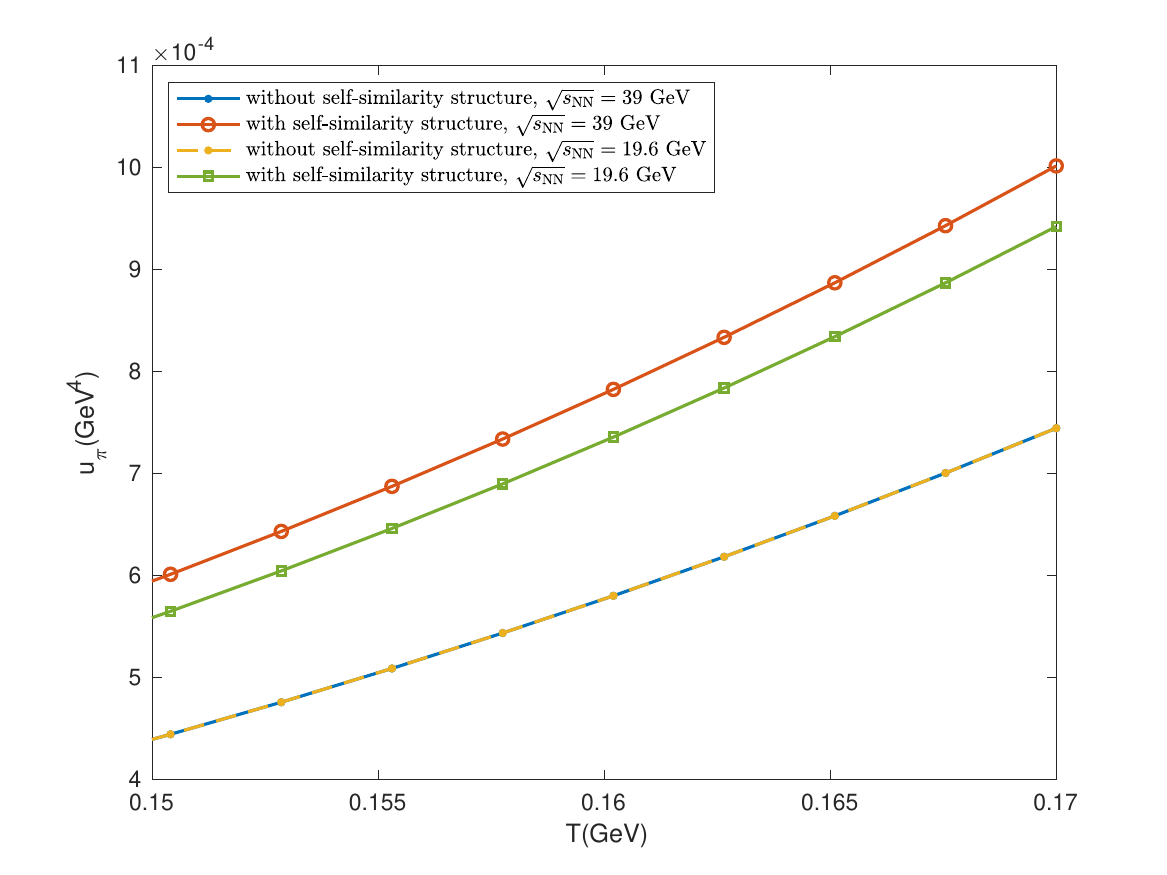} \label{Fig06a}}
\subfloat [Pressure]{\includegraphics[width=0.3\linewidth]{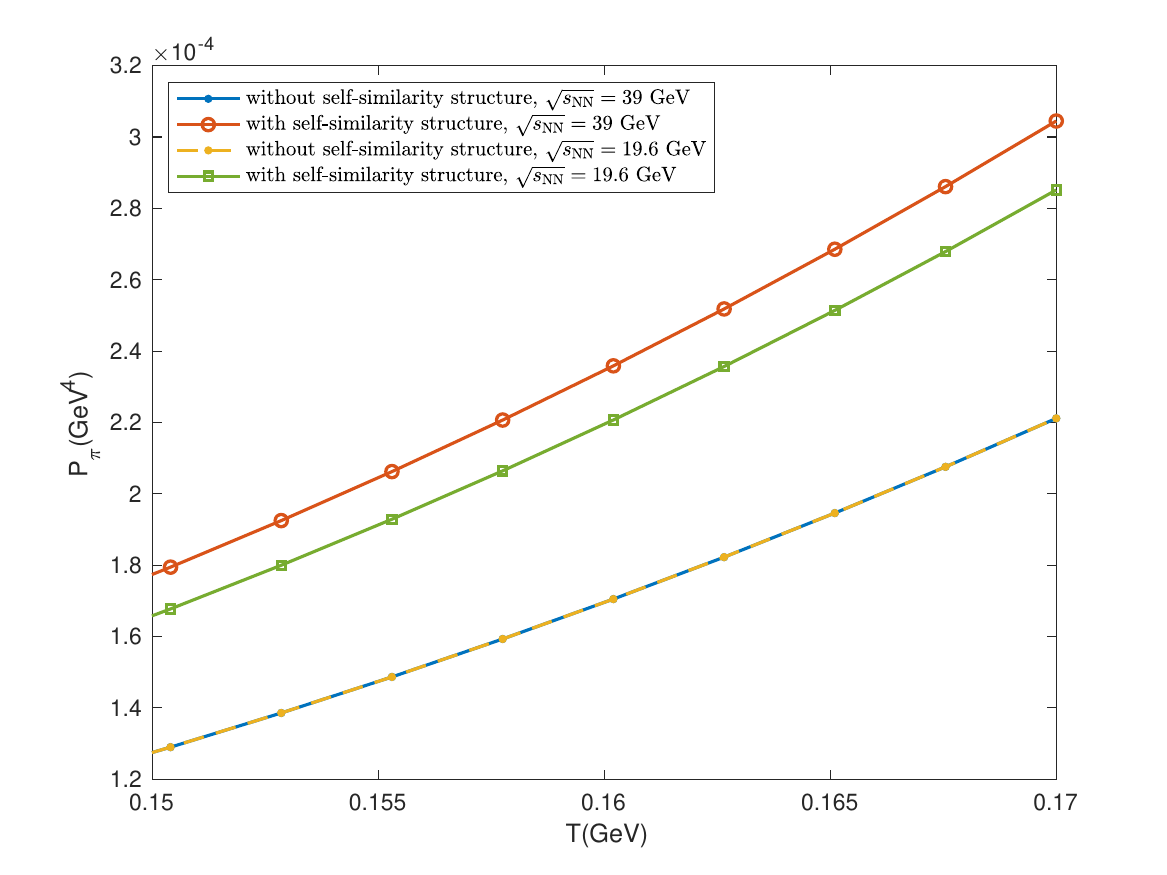} \label{Fig06b}}
\subfloat [Entropy density]{\includegraphics[width=0.3\linewidth]{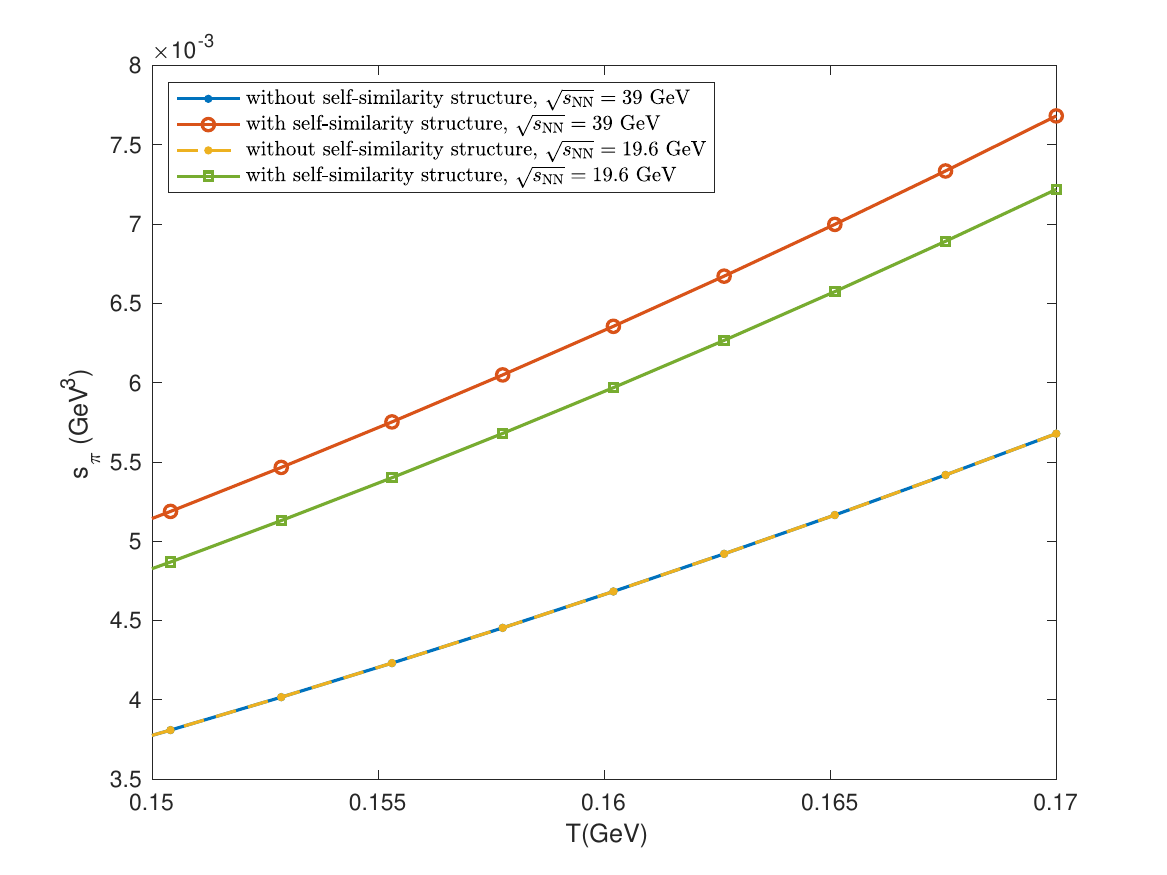} \label{Fig06c}}
\caption{Thermodynamic quantities of pion gas with and without the self-similarity structure influence in Au+Au collisions at $\sqrt{s_{\text{NN}}} = 39, 19.6 \, \text{GeV} $.}
\label{Fig06}  
\end{figure}

We take $q_{1} = 1.03595, 1.04482$ in Au+Au collisions at $\sqrt{s_{\text{NN}}} = 19.6, 39 \, \text{GeV} $ listed in table \ref{TableI} as an example to study the influence of self-similarity structure, caused by resonance, quantum correlation and interaction respectively. The self-similarity structure influences most when approaching the transition temperature. We compare the quantities of pion gas with and without the influence of self-similarity structure just below the transition temperature as shown in figure \ref{Fig06}. Depicted as in figure \ref{Fig06}, it can be found that the energy density, entropy density and pressure without the influence at $\sqrt{s_{\text{NN}}} = 19.6, 39 \, \text{GeV} $ are the same respectively. It is known that without the influence of self-similarity structure ($q=1$), there is the same particle number distribution as Bose-Einstein distribution at different collision energies, leading to the same physical properties. Shown in figure \ref{Fig06a} is the energy density with and without the self-similarity structure at $\sqrt{s_{\text{NN}}} = 19.6, 39 \, \text{GeV} $ collision energy. It can be seen that the energy density increases with the temperature. At fixed temperature for the same collision energy, the energy density under the influence of self-similarity is larger than that without the influence. This is because the interaction between $\pi$ meson causes self-similarity structure leads to the increase of the energy density. At fixed temperature for the different collision energies, the energy density under the influence of self-similarity structure at  $\sqrt{s_{\text{NN}}} = 39 \, \text{GeV} $ (red trace) is larger than that at $\sqrt{s_{\text{NN}}} = 19.6 \, \text{GeV} $ (green trace), illustrating that there is stronger influence of self-similarity structure on pion gas in larger collision energy.

Shown in figure \ref{Fig06b} is the pressure with and without self-similarity structure influence as a function of temperature. It can be seen that the pressure exhibits the same increasing trend as the energy density. This is because at fixed temperature, the larger energy density under the influence of self-similarity structure induces larger pressure. Shown in figure \ref{Fig06c} is the entropy density with and without self-similarity structure as a function of temperature. It shows that the entropy density increases with the temperature. At fixed temperature for the same collision, the entropy density with self-similarity structure influence is larger than that without the influence. This is because the interaction potential between pions, one of the factors contributing to the self-similarity structure, increases the microstates. Consequently, the entropy density increases. At fixed temperature for $\sqrt{s_{\text{NN}}} = 19.6, 39 \, \text{GeV} $, with the influence of self-similarity structure, the entropy density at $\sqrt{s_{\text{NN}}} = 39 \, \text{GeV} $ (red trace) is larger than that at  $\sqrt{s_{\text{NN}}} = 19.6\, \text{GeV} $ (green trace). It indicates a strengthened interaction between $\pi$ mesons with increasing collision energy, leading to more two-body self-similarity structure. When the influence of self-similarity structure increases, the microstates increase and it also leads to the increase of entropy density.

With the particle number distribution of $\pi$, in terms of the transverse momentum $p_T$ and the rapidity $y$, we can derive the transverse momentum distribution function as
\begin{equation}
\frac{d^{2} N}{2\pi p_{T} dp_{T} dy} = \frac{g V_{lab} m_{T} cosh y}{(2 \pi)^{3}}\frac{1}{ [ 1 + (q_{1} -1) \beta m_{T} cosh y]^{\frac{q_{1}}{q_{1} -1}} -1},
\label{pt}
\end{equation}
$p_{T}$ is the transverse momentum in the lab frame, and $m_{T} = \sqrt{m^{2} + p_{T}^{2}}$ is the transverse mass of pion. $V_{lab}$ is the motion volume of $\pi$ meson, related to the volume in center-of-mass frame $V$. Here is the Lorentz transformation $V_{lab} = \gamma V$, $\gamma$ is the Lorentz factor. In equation (\ref{pt}), $q_{1}$ can be solved through TBFM in meson aspect.

\begin{figure}[htbp]
\centering
\includegraphics[scale=0.4]{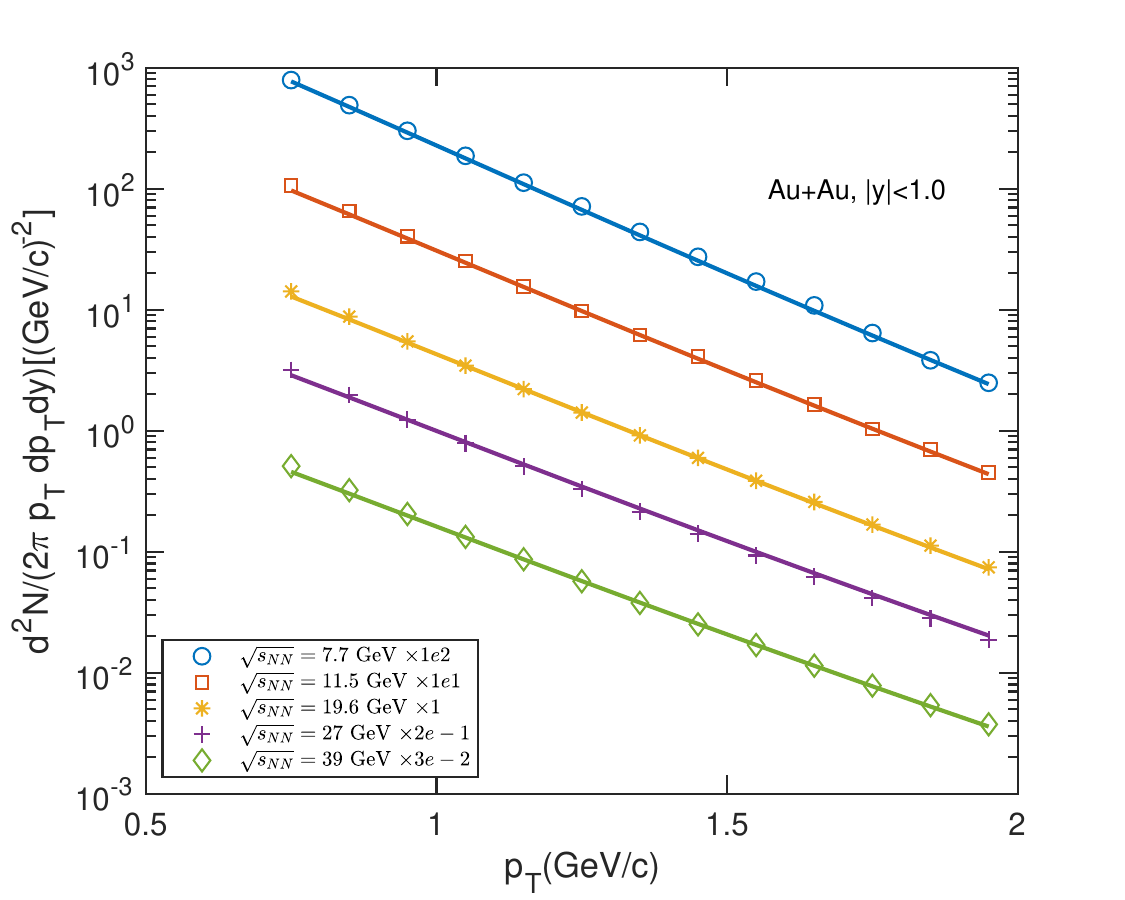}
\caption{Transverse momentum spectra of $\pi^{+}$ meson in Au+Au collisions at $\sqrt{s_{\text{NN}}} = 7.7, 11.5, 19.6, 27, 39 \text{\ GeV} $ for $0-5\%$ centrality, in mid-rapidity $|y|<0.1$. The experimental data are from STAR\cite{64}.}
\label{Fig07}
\end{figure}

Substituting the value of $q_{1}$ from TBFM into equation (\ref{pt}), we can derive the $p_{T}$ spectrum of $\pi$ mesons in Au+Au collisions at $\sqrt{s_{\text{NN}}} = 7.7, 11.5, 19.6, 27, 39 \text{\ GeV} $ for $0-5\%$ centrality, and compare the results with the experimental data shown in figure \ref{Fig07}. As depicted in figure \ref{Fig07}, these traces represent the transverse momentum distribution under the influence of the two-body self-similarity structure for pion gas produced in the collisions at $\sqrt{s_{\text{NN}}} = 7.7, 11.5, 19.6, 27, 39 \text{\ GeV}$, and the colorful dots  are the experimental transverse momentum distribution of these collisions. It can be seen that the theoretical results from TBFM have a good agreement with the experimental data, illustrating that the self-similarity structure exists in the intermediate energy collisions, and influences the distribution of the pion gas. If the distribution of the pion gas is altered by the influence of the self-similarity structure, the phase transition will also be impacted when considering the influence of the self-similarity structure. 

\subsection{QCD phase diagram under the influence of QGP boundary effect and self-similarity structure of pions}
\begin{figure}[htbp]
\centering
\includegraphics[scale=0.4]{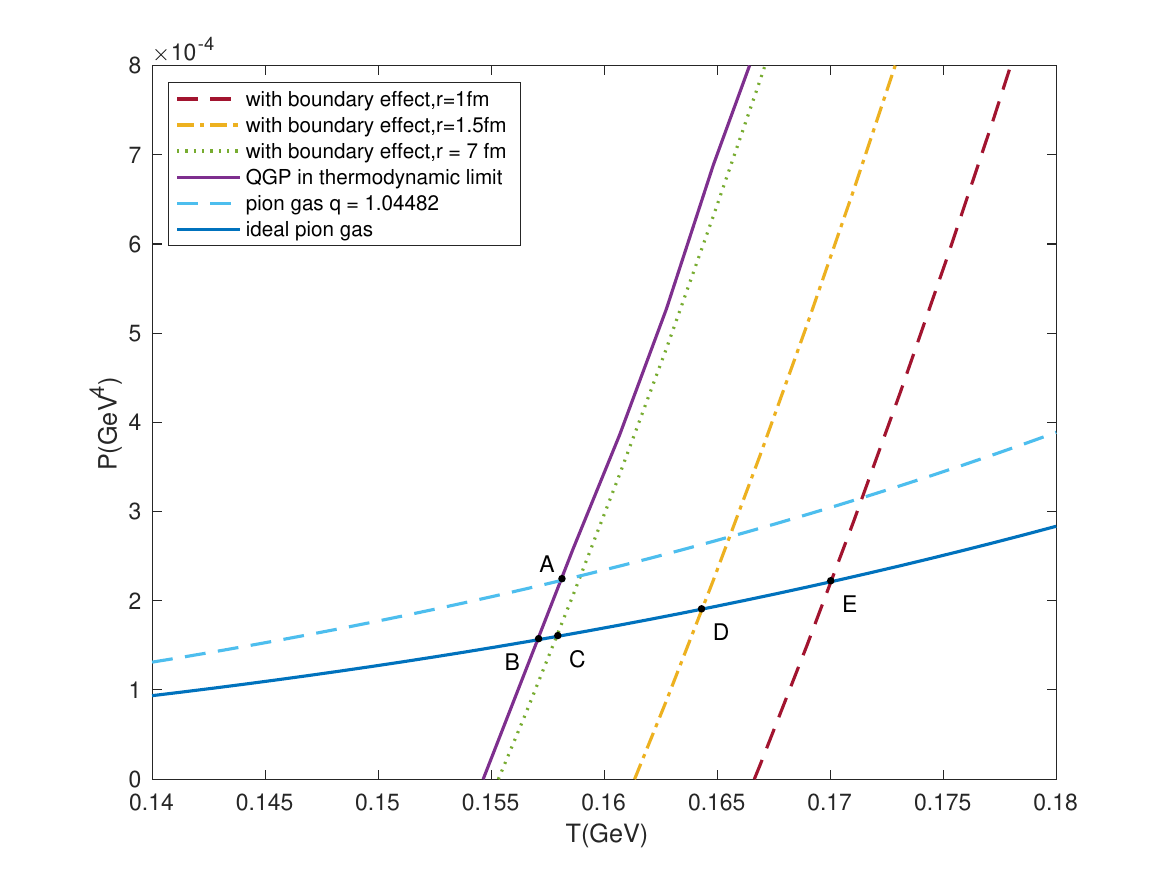}
\caption{The pressure in hadron phase in Au+Au collisions at $\sqrt{s_{\text{NN}}} = 39 \text{\ GeV} $and that of QGP droplets at radius $r = 1, 1.5, 7\text{\ fm}$.}
\label{Fig08}
\end{figure}

\begin{figure}[htbp]
\centering
\includegraphics[width = 0.45\linewidth]{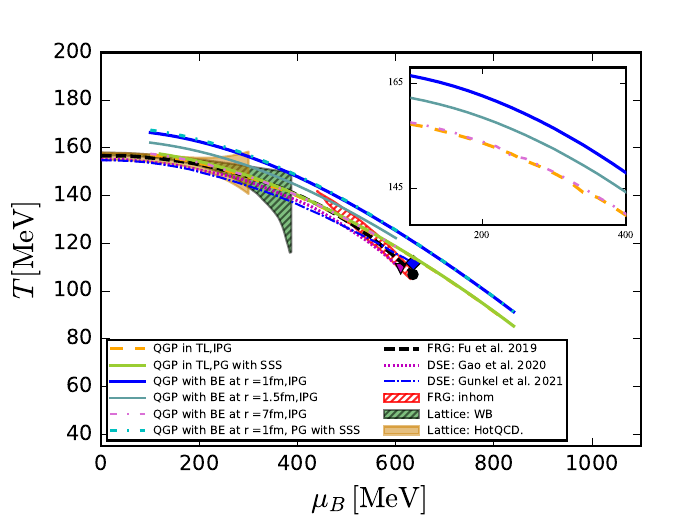}
\caption{The phase diagram with considering (1) QGP in thermodynamic limit (TL) and ideal pion gas (IPG), (2) QGP in thermodynamic limit and pion gas (PG) with the influence of self-similarity structure (SSS), (3)(4)(5) QGP droplets with the boundary effect (BE) at radius  $r = 1 \text{\ fm},1.5 \text{\ fm},7\text{\ fm}$ and ideal pion gas respectively, (6) QGP droplet with the boundary effect (BE) at radius  $r = 1 \text{\ fm}$ and pion gas with the influence of self-similarity structure (SSS). We also list the results from fRG model\cite{65}, DSE\cite{66,67} and lattice QCD\cite{47,48} for comparison.}
\label{Fig09}
\end{figure}

In the heavy-ion collisions, Quark-Gluon Plasma (QGP) undergoes a phase transition and turns into hadrons. As discussed in the above, the properties of QGP are influenced by the boundary effect, and that of hadrons are affected by the self-similarity structure. The influence on the two phases will surely impact the phase transition temperature. Based on the equilibrium condition that $T_{\text{QGP}} = T_{\pi}, P_{\text{QGP}} = P_{\pi}, \mu_{\text{QGP}} = \mu_{\pi}$, with the QGP pressure in equation (\ref{pinbag}) and hadron pressure in equation (\ref{pionp}), the phase transition temperature can be determined. Shown in figure \ref{Fig08} is the pressure of QGP phase and hadron phase in Au+Au collision at $\sqrt{s_{\text{NN}}} = 39 \text{\ GeV}$ with baryon chemical potential $\mu = 0.112 \text{\ GeV}$. The traces in left part correspond to the pressure in hadron phase, those in the right part correspond to the QGP phase, and the intersections correspond to the phase transition points. We also calculate the phase transition temperature at different fixed chemical potential and different collision energies as shown in figure \ref{Fig09}, which displays the phase diagrams in heavy-ion collisions. The details of the figure description and physical analysis are in the following.

In order to study the individual influence of QGP boundary effect and self-similarity structure of pions, we firstly neglect the self-similarity structure on pions and treat pions as the ideal gas through setting $q_{1} =1$ ( blue solid trace) to study the QGP boundary effect. As shown in figure \ref{Fig08}, the blue solid trace intersects with the trace of the pressure of QGP phase in thermodynamic limit at point B, and intersects with these traces under QGP boundary effect at QGP radius $r = 1 \text{\ fm},1.5 \text{\ fm},7\text{\ fm}$ at point C, D, E, respectively. It is observed that the phase transition temperature increases due to the boundary effect and obey the rule $T_{B} < T_{C}<T_{D} <T_{E}$. This increase is attributed to the reduction of QGP pressure due to the boundary effect. In order to reach the phase transition pressure, the phase transition temperature should be increased. The smaller the QGP radius is, the stronger the boundary effect is, leading to a greater increase in the phase transition temperature.

We study not only the phase transition temperature at collision energy $\sqrt{s_{\text{NN}}} = 39 \text{\ GeV}$, $\mu = 0.112 \text{\ GeV}$, but also the temperature at different collision energies and different fixed baryon chemical potentials. As shown in figure \ref{Fig09}, the blue solid trace, the cadetblue solid trace and the pink dotted trace correspond to the phase diagrams with QGP radius $r = 1 \text{\ fm},1.5 \text{\ fm},7\text{\ fm}$, respectively. A detailed presentation of three traces is inset in the upper right part of figure \ref{Fig09}. It can be seen that at different chemical potentials, the trend of the phase transition temperature under the influence of boundary effect is the same as that in $\sqrt{s_{\text{NN}}} = 39 \text{\ GeV}$ case, that $T_{PT, r= 7\text{fm}}<T_{PT, r= 1.5\text{fm}}<T_{PT, r= 1\text{fm}}$. This demonstrates that at different collision energies, the boundary effect also increases the transition temperature. The influence becomes more pronounced with decreasing the QGP droplet radius.

Secondly, we consider QGP in thermodynamic limit and focus on the self-similarity in hadron phase to study the influence of resonance, interaction and quantum correlation on phase transition temperature. As shown in figure \ref{Fig08}, the intersection A of the blue dashed trace in hadron phase and the purple solid trace in QGP phase corresponds to the phase transition point under the influence of self-similarity structure in hadron phase. It can be seen that the phase transition temperature $T_{A}$ under the influence of self-similarity structure is larger than that without the influence $T_{B}$. This is because the self-similarity structure increases the pressure in hadron phase. In order to have the two traces of the pressure in the two phases to intersect, the phase transition pressure should be elevated, so that the phase transition temperature increases. We also compute the phase transition temperature under the influence of self-similarity structure in hadron phase at different collision energies and different chemical potentials based on equation (\ref{fit}) as shown the green solid trace in figure \ref{Fig09}. It can be seen that this trace is a bit above the ideal gas (orange dashed trace) in low and mid chemical potential region, suggesting that self-similarity structure increases the transition temperature in this region. In high chemical potential region, the transition temperature is approximately the same as that for ideal pion gas. This is consistent with the prediction of TBFM discussed in section~\href{sec2.2}{2.2}, that the influence of the self-similarity structure decreases with increasing the chemical potential. In the higher chemical potential region, the collision energy is smaller, leading to smaller collective flow velocity, so that the resonance effect is smaller, and the influence of self-similarity structure diminishes correspondingly.

 Lastly, we take the boundary effect of the QGP droplets and the influence of two-body similarity structure on the pion gas both into account, and explore the phase transition in the intermediate and high baryon chemical potential region ($0.4\sim 0.85 \text{\ GeV}$), which can be produced in the future HIAF. Illustrated in figure \ref{Fig09}, the phase transition of QGP in thermodynamic limit (orange dashed trace), the phase transition with boundary effect of QGP droplets at radius $r = 1 \text{\ fm}$ (blue solid trace), and that with boundary effect of QGP droplet and the two-body similarity structure influence on pion gas (cyan dotted trace) are plotted. And we also include the phase transition results from fRG (functional renormalization group) approach\cite{65}, DSE (Dyson-Schwinger equations) model\cite{66,67}, lattice QCD\cite{47,48} in finite chemical potential region for comparison. 

As shown in figure \ref{Fig09}, the cyan dotted trace (under the influence of boundary effect in QGP phase and self-similarity structure in hadron phase) is a bit above the blue solid trace (only under the influence of boundary effect in QGP phase) in low and mid chemical potential region, and approximately coincides with the blue solid trace. The reason is explained in the above that the influence of self-similarity structure decreases with increasing chemical potential. It is also found that the cyan dotted trace is much higher than the green solid trace (only under the influence of self-similarity structure) and orange dashed trace (QGP in thermodynamic limit, pion ideal gas), indicating that the boundary effect with small QGP radius increases the phase transition temperature more than the self-similarity structure in hadron phase. This is because the pressure under the influence of boundary effect in QGP phase increases faster than that under the influence of self-similarity structure in hadron phase. This faster increase of pressure under the influence of boundary effect induces larger increase of phase transition temperature.

\section{Conclusion}
We study the curved boundary effect on QGP droplet and the self-similarity structure effect on pion gas in intermediate and low energy heavy-ion collisions. In QGP phase, we use the modified MIT bag model with MRE method to study the boundary effect on QGP droplet. At fixed temperature, it is found that the energy density, entropy density, and pressure with the boundary effect are smaller than those without this effect. The boundary curves more, the energy density, entropy density and pressure decrease more.

In hadron phase, we consider the resonance, strong interaction, and quantum effects, which induce the self-similarity of  $\pi$-$\pi$ two-meson state and $u\bar{d}$ two-quark state of pion. We develop a Two-Body Fractal Model (TBFM) to investigate the influence of self-similarity structure. In the framework of TBFM, we introduce the modification factor $q_{1}$ to represent the self-similarity structure influence on the $\pi$ meson, and the escort parameter $q_{2}$ to characterize the self-similarity structure influence on the quarks. We derive the probability and Tsallis entropy of $\pi$ meson from quark and meson aspects, and solve the corresponding equations to obtain the values of $q_{1}$, $q_{2}$. The result $q_{1} > 1$ suggests that the $\pi$ gas is influenced by the $\pi$-$\pi$ two-body structures arising from the resonance, the interaction, and quantum correlation effects. Furthermore, the value of $q_{1}$ decreases with decreasing the temperature, prompting us to analyse the self-similarity structure influence near the phase transition temperature. We simulate the temperature variation with the chemical potential from lattice QCD and obtain $q_{1}$ as a function of chemical potential $\mu$. It is found that $q_{1}$ decreases as the chemical potential increases. This is consistent with the trend where decreasing the collision energy and increasing the chemical potential lead to the decrease of the collective flow velocity, and subsequently decrease the influence of self-similarity structure. It is predicted that $q_{1} = 1.0208 \sim 1.0224 $ in future HIAF energy region $\sqrt{s_{\text{NN}}} = 2.2 \sim 4.5 \text{\ GeV}$. We compute the $\pi$ meson distribution with the self-similarity structure influence, and derive the energy density, entropy density and pressure of $\pi$ meson. It is found that the energy, entropy and pressure of $\pi$ meson increase compared to the results in thermodynamic limit. Substituting the $q_{1}$ into the non-extensive transverse momentum spectrum, it is observed to align well with the experimental data of Au+Au collision at $\sqrt{s_{\text{NN}}} = 7.7, 11.5, 19.6, 27, 39 \text{\ GeV} $ .

Finally, we explore the hadron-QGP phase transition under the influences of the boundary effect in QGP and two-body self-similarity structure in pion gas. Considering the finite size QGP droplet, and utilizing the results of TBFM in intermediate chemical potential region, we derive the phase diagram with these effects, and compare the result to that without these influences. Considering the boundary effect of QGP and treating hadron phase as the ideal pion gas, we find that the phase transition temperature increases with the boundary effect. The smaller QGP droplet and stronger boundary effect are, the more the phase transition temperature increases. While considering QGP in thermodynamic limit and pion gas influenced by the self-similarity structure, the phase transition temperature increases slightly with the influence in low and mid chemical potential, and remains approximately the same as that without the influence in high chemical potential. This suggests that there is the smaller collective flow velocity in lower energy collisions, causing the smaller resonance effect and the diminished influence of self-similarity structure. Considering both the boundary effect of QGP and two-body self-similarity structure effect of pion gas, our analysis reveals that the phase transition temperature is larger compared with that without the two effects, and increases slightly compared to that with only boundary effect of QGP. This signifies the notable influence of boundary effect and surface tension in the high chemical potential region where the small QGP droplet can be produced. Instead, the influence of the self-similarity structure diminishes as the collective flow velocity decreases in this region. The result is instructive to the future low-energy experiments that aimed at observing the physics of the intermediate and high chemical potential region.

\vspace*{2mm}
\vspace*{-1mm}
\begin{small}\baselineskip=10pt\itemsep-2pt

\begin{thebibliography}{10}

\bibitem{1}
Singh~C~P~1993 
\newblock Signals of quark-gluon plasma
\newblock {\em Phys. Rep.} 
\href{https://doi.org/10.1016/0370-1573(93)90172-A}
{\textbf{236} 147}

\bibitem{2}
Satz~H~2000
\newblock Colour deconfinement in nuclear collisions 
\newblock {\em Rep. Prog. Phys.} 
\href{https://dx.doi.org/10.1088/0034-4885/63/9/203}{\textbf{63} 1511}

\bibitem{3}
 Srivastava~P~K, Tiwari~S~K~and~Singh~C~P~2010
\newblock {QCD} critical point in a quasiparticle model
\newblock {\em Phys. Rev. D} \href{https://link.aps.org/doi/10.1103/PhysRevD.82.014023} {\textbf{82} 014023}


\bibitem{4}
Arsene~I~\textit{et al}~(BRAHMS)~2005
\newblock Quark–gluon plasma and color glass condensate at RHIC? The perspective from the BRAHMS experiment
\newblock {\em Nucl. Phys. A}
\href{https://doi.org/10.1016/j.nuclphysa.2005.02.130} {\textbf{757} 1}

\bibitem{5}
Back~B~B~\textit{et al}~(PHOBOS)~2005
\newblock The {PHOBOS} perspective on discoveries at {RHIC}
\newblock {\em Nucl.Phys.A} 
\href{https://doi.org/10.1016/j.nuclphysa.2005.03.084}  {\textbf{757} 28}


\bibitem{6}
Adams ~J \textit{et al} (STAR) 2005
\newblock Experimental and theoretical challenges in the search for the {Q}uark–{G}luon {P}lasma: The {STAR} {Collaboration's} critical assessment of the evidence from {RHIC} collisions
\newblock {\em Nucl.Phys.A}
\href{https://doi.org/10.1016/j.nuclphysa.2005.03.085}{\textbf{757} 102}


\bibitem{7}
Mohanty~B~2009
\newblock {QCD Phase Diagram: Phase Transition, Critical Point and Fluctuations}
\newblock {\em Nucl.Phys.A}
\href{https://doi.org/10.1016/j.nuclphysa.2009.10.132}{\textbf{830} 899}

\bibitem{8}
An~X~\textit{et al}~2022
\newblock The {BEST} framework for the search for the {QCD} critical point and the chiral magnetic effect
\newblock {\em Nucl.Phys.A}
\href{https://doi.org/10.1016/j.nuclphysa.2021.122343}{\textbf{1017} 122343}

\bibitem{9}
Karthein~J~M~\textit{et al}~2021
\newblock {Strangeness-neutral equation of state for QCD with a critical point}
\newblock {\em Eur. Phys. J. Plus} 
\href{https://doi.org/10.1140/epjp/s13360-021-01615-5}{\textbf{136} 621}

\bibitem{10}
Odyniec~G~(STAR)~2019
\newblock {Beam Energy Scan Program at RHIC (BES I and BES II) \textendash{} probing {QCD} phase diagram with heavy-ion collisions}
\newblock {\em PoS}
\href{https://doi.org/10.22323/1.347.0151}{\textbf{CORFU2018} 151 }

\bibitem{11}
Bzdak~A, Esumi~S, Koch~V, Liao~J-F, Stephanov~M~and~Xu~N~2020
\newblock {Mapping the phases of {Quantum Chromodynamics with Beam Energy Scan}}.
\newblock {\em Phys. Rept.} 
\href{https://doi.org/10.1016/j.physrep.2020.01.005}{\textbf{853} 1}

\bibitem{12}
Li~P-C~\textit{et al}~2023
\newblock Effects of a phase transition on two-pion interferometry in heavy ion collisions at {$\sqrt{s_{NN}}= 2.4- 7.7$ {GeV}}
\newblock {\em Sci. China Phys. Mech. Astron.}
\href{https://doi.org/10.1007/s11433-022-2041-8}{\textbf{66} 232011} 

\bibitem{13}
Luo~X-F, Shi~S-S, Xu~N~and~Zhang~Y~F~2020
\newblock A study of the properties of the {QCD} phase diagram in high-energy nuclear collisions
\newblock {\em Particles}
\href{ https://doi.org/10.3390/particles3020022}{\textbf{3} 278}

\bibitem{14}
Bringoltz~B~and~Teper~M~2005
\newblock {The pressure of the $SU(N)$ lattice gauge theory at large $N$}
\newblock {\em Phys. Lett. B}
\href{https://doi.org/10.1016/j.physletb.2005.08.127}{\textbf{628} 113}

\bibitem{15}
Callaway~D~J~E and Rahman~A~1982
\newblock {Microcanonical Ensemble Formulation of Lattice Gauge Theory}
\newblock {\em Phys. Rev. Lett.}
\href{https://doi.org/10.1103/PhysRevLett.49.613}{\textbf{49} 613}

\bibitem{16}
Geloun~J~B, Martini~R~and~Oriti~D~2016
\newblock Functional renormalization group analysis of tensorial group field theories on{ ${\mathbb{R}}^{d}$}
\newblock {\em Phys. Rev. D}
\href{https://doi.org/10.1103/PhysRevD.94.024017}{\textbf{94} 024017}

\bibitem{17}
Aad~G~\textit{et al}~(ATLAS)~2010
\newblock Observation of a centrality-dependent dijet asymmetry in lead-lead collisions at {$\sqrt{{s}_{NN}}=2.76\text{ }\mathrm{TeV}$} with the {ATLAS Detector at the LHC}
\newblock {\em Phys. Rev. Lett.}
\href{https://doi.org/10.1103/PhysRevLett.105.252303}{\textbf{105} 252303}

\bibitem{18}
Kolbé~I, Horowitz~W~A~and~Mogliacci~S~2019
\newblock Small system corrections to thermal field theory and p{QCD} energy loss
\newblock {\em J. Phys.: Conf. Ser.}
\href{https://dx.doi.org/10.1088/1742-6596/1271/1/012019}{\textbf{1271} 012019}

\bibitem{19}
Chatrchyan~S~\textit{et al}~(CMS)~2011
\newblock Observation and studies of jet quenching in {PbPb} collisions at {$\sqrt{{s}_{NN}}=2.76\text{ }\mathrm{TeV}$}
\newblock {\em Phys. Rev. C}
\href{https://doi.org/10.1103/PhysRevC.84.024906}{\textbf{84} 024906}

\bibitem{20}
Aamodt~K~\textit{et al}~(ALICE)~2011
\newblock Suppression of charged particle production at large transverse momentum in central {Pb–Pb} collisions at {$\sqrt{{s}_{NN}}=2.76\text{ }\mathrm{TeV}$}
\newblock {\em Phys. Lett. B}
\href{https://doi.org/10.1016/j.physletb.2010.12.020}{\textbf{696} 30}


\bibitem{21}
Venugopalan~R~and~Prakash~M~1992
\newblock Thermal properties of interacting hadrons
\newblock {\em Nucl.Phys.A}
\href{https://doi.org/10.1016/0375-9474(92)90005-5}{\textbf{546} 718}

\bibitem{22}
Deur~A, Brodsky~S~J~and~{de Téramond}~G~F~2016
\newblock {The QCD running coupling}
\newblock {\em Prog. Part. Nucl. Phys.}
\href{https://doi.org/10.1016/j.ppnp.2016.04.003}{\textbf{90} 1}

\bibitem{23}
Sahoo~R~2019
\newblock {Possible formation of {QGP}-droplets in proton-proton collisions at the {CERN Large Hadron Collider}}
\newblock {\em AAPPS Bull.}
\href{https://doi.org/10.22661/AAPPSBL.2019.29.4.16}{\textbf{29} 16}

\bibitem{24}
Niida~T~and~Miake~Y~2021
\newblock {Signatures of QGP at RHIC and the LHC}
\newblock {\em AAPPS Bull.}
\href{https://doi.org/10.1007/s43673-021-00014-3}{\textbf{31} 12}

\bibitem{25}
Loizides~C~2016
\newblock Experimental overview on small collision systems at the {LHC}
\newblock {\em Nucl.Phys.A}
\href{https://doi.org/10.1016/j.nuclphysa.2016.04.022}{\textbf{956} 200}

\bibitem{26}
Wong~C-Y~and~Zhang~W-N~2007
\newblock {Single-Event Handbury-Brown-Twiss interferometry}
\newblock {\em Int. J. Mod. Phys. E}
\href{https://doi.org/10.1142/S0218301307009245}{\textbf{16} 3271}

\bibitem{27}
Shneider~M~N~and~Pekker~M~2019
\newblock Surface tension of small bubbles and droplets and the cavitation threshold
\newblock {arXiv:}\href{https://doi.org/10.48550/arXiv.1901.04329}{1901.04329}

\bibitem{28}
Danielewicz~P~and~Odyniec~G~1985
\newblock Transverse momentum analysis of collective motion in relativistic nuclear collisions
\newblock {\em Phys. Lett. B}
\href{https://doi.org/10.1016/0370-2693(85)91535-7}{\textbf{157} 146}

\bibitem{29}
Gustafsson~H~A~\textit{et~al}~1984
\newblock Collective flow observed in relativistic nuclear collisions
\newblock {\em Phys. Rev. Lett.}
\href{https://doi.org/10.1103/PhysRevLett.52.1590}{\textbf{52} 1590}

\bibitem{30}
Wiranata~A, Koch~V, Prakash~M~and~Wang~X-N~2014
\newblock Shear viscosity of a multi-component hadronic system
\newblock {\em J. Phys.: Conf. Ser.}
\href{https://dx.doi.org/10.1088/1742-6596/509/1/012049}{\textbf{509} 012049}


\bibitem{31}
Zachariasen~F and Zemach~C~1962
\newblock  Pion resonances
\newblock {\em Phys. Rev.}
\href{https://doi.org/10.1103/PhysRev.128.849}{\textbf{128} 849}

\bibitem{32}
Hansson~T~H~and~Jaffe~R~L~1983
\newblock {\em The multiple reflection expansion for confined scalar, Dirac, and gauge fields}
\newblock{\em Ann. Phys.}
\href{https://doi.org/10.1016/0003-4916(83)90319-6}{\textbf{151} 204}

\bibitem{33}
Ding H-Q~\textit{et~al}~2023
\newblock The spectrum of low-$p_{T} {J}/\psi$ in heavy-ion collisions in a statistical two-body fractal model
\newblock {\em Entropy}
\href{ https://doi.org/10.3390/e25121655}{\textbf{25} 1655 }

\bibitem{34}
Becattini~B~2014
\newblock The Quark Gluon Plasma and relativistic heavy ion collisions in the LHC era
\newblock{\em J. Phys.: Conf. Ser.} 
\href{https://dx.doi.org/10.1088/1742-6596/527/1/012012}{\textbf{527} 012012}

\bibitem{35}
Moreau~P~\textit{et al}~2019
\newblock Exploring the partonic phase at finite chemical potential within an extended off-shell transport approach
\newblock {\em Phys. Rev. C}
\href{https://doi.org/10.1103/PhysRevC.100.014911}{\textbf{100} 014911}

\bibitem{36}
Shen~C~and~Alzhrani~S~2020
\newblock {Collision-geometry-based 3D initial condition for relativistic heavy-ion collisions}
\newblock {\em Phys. Rev. C}
\href{https://doi.org/10.1103/PhysRevC.102.014909}{\textbf{102} 014909}

\bibitem{37}
Adam~J~\textit{et~al}~(STAR)~2020
\newblock {Strange hadron production in Au+Au collisions at $\sqrt{s_{_{\mathrm{NN}}}}$ = 7.7, 11.5, 19.6, 27, and 39 GeV}
\newblock {\em Phys. Rev. C}
\href{https://doi.org/10.1103/PhysRevC.102.034909}{\textbf{102} 034909}

\bibitem{38}
Albacete~J~L, Guerrero-Rodr\'iguez~P~and~Marquet~C~2019
\newblock Initial correlations of the Glasma energy-momentum tensor
\newblock {\em J. High Energ. Phys.}
\href{https://doi.org/10.1007/JHEP01(2019)073}{\textbf{2019} 73}

\bibitem{39}
Odyniec~G~1999
\newblock Statistical and dynamical fluctuations in heavy ion collisions: Role of conservation laws in event-by-event analysis
\newblock {\em Acta Phys. Polon. B}
\href{https://www.osti.gov/biblio/1407016}{\textbf{30} 385}

\bibitem{40}
Chodos~A, Jaffe~R~L, Johnson~K, Thorn~C~B~and~Weisskopf~V~F~1974
\newblock New extended model of hadrons
\newblock {\em Phys. Rev. D}
\href{https://doi.org/10.1103/PhysRevD.9.3471}{\textbf{9} 3471}

\bibitem{41}
Ramanathan~R, Gupta~K~K, Jha~A~K~and~Singh~S~S~2007
\newblock {The interfacial surface tension of a quark-gluon plasma fireball in a hadronic medium}
\newblock {\em Pramana}
\href{https://doi.org/10.1007/s12043-007-0075-8}{\textbf{68} 757}

\bibitem{42}
Madsen~J~1993
\newblock Curvature contribution to the mass of strangelets
\newblock {\em Phys. Rev. Lett.}
\href{https://doi.org/10.1103/PhysRevLett.70.391}{\textbf{70} 391}

\bibitem{43}
Madsen~J~1994
\newblock Shell model versus liquid drop model for strangelets
\newblock {\em Phys. Rev. D}
\href{https://doi.org/10.1103/PhysRevD.50.3328}{\textbf{50} 3328}

\bibitem{44}
Balian~R~and~Bloch~C~1970
\newblock Distribution of eigenfrequencies for the wave equation in a finite domain
\newblock {\em Ann. Phys.}
\href{https://doi.org/10.1016/0003-4916(70)90497-5}{\textbf{60} 401}

\bibitem{45}
Patra~B~K~and~Singh~C~P~1996
\newblock Temperature and baryon-chemical-potential-dependent bag pressure for a deconfining phase transition
\newblock {\em Phys. Rev. D}
\href{https://doi.org/10.1103/PhysRevD.54.3551}{\textbf{54} 3551}

\bibitem{46}
Gao~S, Wang~E-K~and~Li~J-R~1992
\newblock Bag constant and deconfinement phase transition in a nontopological soliton model
\newblock {\em Phys. Rev. D}
\href{https://doi.org/10.1103/PhysRevD.46.3211}{\textbf{46} 3211}

\bibitem{47}
Bazavov~A~\textit{et~al}~(HotQCD)~2019
\newblock Chiral crossover in {QCD} at zero and non-zero chemical potentials
\newblock {\em Phys. Lett. B}
\href{https://doi.org/10.1016/j.physletb.2019.05.013}{\textbf{795} 15}

\bibitem{48}
Bellwied~R~\textit{et~al}~2015
\newblock The {QCD} phase diagram from analytic continuation
\newblock {\em Phys. Lett. B}
\href{https://doi.org/10.1016/j.physletb.2015.11.011}{\textbf{751} 559}

\bibitem{49}
Hagedorn~R~1973
\newblock Thermodynamics of strong interactions
\newblock{\em Cargese Lect. Phys.}
\href{https://cds.cern.ch/record/2010310}{\textbf{6} 643}

\bibitem{50}
Crater~H~W, Yoon~J-H, and Wong~C-Y~2009
\newblock {Singularity structures in Coulomb-type potentials in two-body Dirac equations of constraint dynamics}
\newblock {\em Phys. Rev. D}
\href{https://doi.org/10.1103/PhysRevD.79.034011}{\textbf{79} 264}

\bibitem{51}
Mandelbrot~B~B~1967
\newblock How long is the coast of {B}ritain? {S}tatistical self-similarity and fractional dimension
\newblock {\em Science}
\href{https://doi.org/10.1126/science.156.3775.636}{\textbf{156} 636}

\bibitem{52}
Mandelbrot~B~B~1986
\newblock { Self-affine fractal sets, {I}: The basic fractal dimensions}
\newblock In {\em Fractals in physics} (Amsterdam:Elsevier)
\href{https://doi.org/10.1016/B978-0-444-86995-1.50004-4}{3-15}

\bibitem{53}
Tsallis~C~1988
\newblock Possible generalization of {Boltzmann}-{Gibbs} statistics
\newblock {\em J. Stat. Phys}
\href{https://doi.org/10.1007/BF01016429}{\textbf{52} 479}

\bibitem{54}
Abe~S~and~Okamoto~Y~2001
\newblock {\em Nonextensive statistical mechanics and its applications} Vols. 560
\newblock (Berlin: Springer Science \& Business Media)

\bibitem{55}
Tsallis~C~2009
\newblock {\em Introduction to nonextensive statistical mechanics: approaching a complex world}, Vols.~1
\newblock (Berlin: Springer)

\bibitem{56}
Feng~X, Jin~L-C~and~Riberdy~M~J~2022
\newblock {Lattice QCD Calculation of the pion mass splitting}
\newblock {\em Phys. Rev. Lett.}
\href{https://doi.org/10.1103/PhysRevLett.128.052003}{\textbf{128} 052003}

\bibitem{57}
Wang~G, Liang~J, Draper~T, Liu~K-F, Yang~Y-B~2021
\newblock Lattice calculation of pion form factors with overlap fermions
\newblock {\em Phys. Rev. D}
\href{https://doi.org/10.1103/PhysRevD.104.074502}{\textbf{104} 07452}

\bibitem{58}
Ubriaco~M~R~1999
\newblock Thermodynamics of boson and fermion systems with fractal distribution functions
\newblock {\em Phys. Rev. E}
\href{https://doi.org/10.1103/PhysRevE.60.165}{\textbf{60} 165}

\bibitem{59}
Büyükkiliç~F~and~Demirham~D~1993
\newblock A fractal approach to entropy and distribution functions
\newblock {\em Phys. Lett. A}
\href{https://doi.org/10.1016/0375-9601(93)91118-O}{\textbf{181} 24}

\bibitem{60}
Rajagopal~A~K, Mendes~R~S~and~Lenzi~E~K~1998
\newblock Quantum statistical mechanics for nonextensive systems: Prediction for possible experimental tests
\newblock {\em Phys. Rev. Lett.}
\href{https://doi.org/10.1103/PhysRevLett.80.3907}{\textbf{80} 3907}

\bibitem{61}
Abe~S~2001
\newblock General pseudoadditivity of composable entropy prescribed by the existence of equilibrium
\newblock {\em Phys. Rev. E}
\href{https://doi.org/10.1103/PhysRevE.63.061105}{\textbf{63} 061105}

\bibitem{62}
Wang~Q~A~2002
\newblock Quantum gas distribution prescribed by factorization hypothesis of probability
\newblock {\em Chaos, Solitons Fractals}
\href{https://doi.org/10.1016/S0960-0779(02)00035-8}{\textbf{14} 765}

\bibitem{63}
Beck~C~2000
\newblock Non-extensive statistical mechanics and particle spectra in elementary interactions
\newblock {\em Physica A}
\href{https://doi.org/10.1016/S0378-4371(00)00354-X}{\textbf{286} 164}

\bibitem{64}
 Adamczyk~L~\textit{et~al}~(STAR)~2017
\newblock {Bulk properties of the medium produced in relativistic heavy-ion collisions from the beam energy scan program}
\newblock {\em Phys. Rev. C}
\href{https://doi.org/10.1103/PhysRevC.96.044904}{\textbf{96} 044904}

\bibitem{65}
Fu~W-J, Pawlowski~J~M~and~Rennecke~F~2020
\newblock {QCD} phase structure at finite temperature and density
\newblock {\em Phys. Rev. D}
\href{https://doi.org/10.1103/PhysRevD.101.054032}{\textbf{101} 054032}

\bibitem{66}
Gao~F~and~Pawlowski~J~M~2021
\newblock {Chiral phase structure and critical end point in QCD}
\newblock {\em Phys. Lett. B}
\href{https://doi.org/10.1016/j.physletb.2021.136584}{\textbf{820} 136584}

\bibitem{67}
Gunkel~P~J~and~Fischer~C~S~2021
\newblock {Locating the critical endpoint of QCD: Mesonic backcoupling effects}
\newblock {\em Phys. Rev. D}
\href{https://doi.org/10.1103/PhysRevD.104.054022}{\textbf{104} 054022}

\end{thebibliography}
\end{small}

\end{document}